\documentclass[twocolumn,amssymb,aps,showpacs]{revtex4}

\usepackage{graphicx}
\usepackage{dcolumn}
\usepackage{bm}
\usepackage{amstext}

\newcommand{\trp}[1]{{#1}}

\begin{document}

\title{Role of slip between a probe particle and a gel in microrheology}
\author{Henry C. Fu}
\email{Henry_Fu@brown.edu}
\affiliation{Division of Engineering, Brown University, Providence, RI 02912}
\author{Vivek B. Shenoy}%
\email{Vivek_Shenoy@brown.edu}
\affiliation{Division of Engineering, Brown University, Providence, RI 02912}
\author{Thomas R. Powers}
\email{Thomas_Powers@brown.edu}

\affiliation{Division of Engineering, Brown University, Providence, RI 02912}


\date{\today}

\begin{abstract}

In the technique of microrheology, macroscopic
rheological parameters as well as information about local structure
are deduced from the behavior of microscopic probe particles under thermal or
active forcing.  Microrheology requires knowledge of the relation
between macroscopic parameters and the force felt by a particle in
response to displacements.  We investigate this response function for
a spherical particle 
using the two-fluid model, in which the gel is represented by a
polymer network coupled to a surrounding solvent via a drag force.  We
obtain an analytic solution for the response function in the limit of
small volume fraction of the polymer network, and neglecting inertial
effects.  We use no-slip boundary conditions for the solvent at the
surface of the sphere.  The boundary condition for the network at the
surface of the sphere is a kinetic friction law, for which 
the tangential stress of the network is proportional to relative
velocity of the network and the sphere.  This boundary condition 
encompasses both no-slip and frictionless boundary conditions as limits.  
Far from the sphere there is no
relative motion between the solvent and network due to the coupling
between them.  However, the different boundary conditions on the solvent and
network tend to produce different far-field motions.  We show that the far field motion and the force on the
sphere are controlled by the solvent boundary conditions at high
frequency and by the network boundary conditions at low frequency.  
At low frequencies compression of the network can also affect the force on the
sphere.  We find the crossover frequencies at which the effects of
sliding of the sphere past the polymer network and compression of the gel become important.  The effects of sliding alone can lead to an
underestimation of moduli by up to 33\%, while the effects of compression alone can lead to an
underestimation of moduli by up to 20\%, and the effects of sliding and
compression combined can lead to an underestimation of moduli by up to 43\%.  
\end{abstract}




\maketitle

Single particle microrheology~\cite{Mason_Weitz1995,Schnurr1997} has been a useful tool for
measuring the material characteristics in situations where traditional
rheometers are difficult to use.  For example, it has been
particularly useful in obtaining rheological measurements when large
sample sizes are difficult to obtain, and when removing samples
from their natural environment may be undesirable, such as in living
cells. 
Traditional rheological measurements typically obtain the
frequency-dependent macroscopic shear modulus $G(\omega)$,
which relates shear stresses to homogeneous shear deformations. 
 To interpret the results of single particle microrheological
measurements, one must know what the response of a particle embedded
in the material of question will be to driving forces (or
equivalently, the force exerted on the particle 
in response to particle displacements).  In practice, the response of
such a particle has been assumed to relate to the macroscopic
shear modulus via 
\begin{equation}
\mathbf{f} = -6 \pi a \mathrm{Re} \left\lbrace G(\omega)  \delta
\mathbf{r} \exp(-i \omega t) \right\rbrace \label{rheo},
\end{equation}
when the particle with radius $a$ oscillates at frequency $\omega$ and
displacement $\mathrm{Re} \lbrace \delta \mathbf{r} \exp(-i \omega t)
\rbrace$.  The modulus $G= G' - i G''$ is in general complex.  In microrheological
experiments, the use of this force response in conjunction with the
fluctuation-dissipation theorem has been called the ``generalized Stokes-Einstein relation''\cite{Mason_Weitz1995}.

Eq.~\ref{rheo}  holds true when the
material may treated using continuum mechanics as a single incompressible
phase with a complex shear modulus and no-slip
boundary conditions between the particle and the material
~\cite{Schnurr1997, Oestreicher1951}.    The generalized Stokes-Einstein relation has been validated experimentally in the test case of
a solution of wormlike micelles~\cite{Buchanan2005}, but in many cases
there are a number of issues that can complicate microrheological
measurements, including compressional effects~\cite{Schnurr1997,
  Levine_Lubensky2001a}, local depletion of the polymer~\cite{Crocker2000, Chen2003}, modification of local properties
via surface chemistry of the particles~\cite{McGrath2000, Valentine2004}, and
violation of no-slip boundary conditions~\cite{Starrs_Bartlett2003}.

Although the technique of 2-particle microrheology~\cite{Crocker2000,Levine_Lubensky2000} ameliorates many of
these problems, it is much more technically demanding than single-particle
microrheology.  Furthermore, fully understanding how all the above effects can
impact 1-particle microrheology allows us to use it to understand
materials and processes in which local heterogeneity plays an
important role.  In this paper we demonstrate how to quantify the
effects arising from violation of the no-slip boundary
condition for both an incompressible and compressible gel.  Previous
studies have treated the response of the medium
to a particle in the case of no-slip boundary
conditions~~\cite{Oestreicher1951, Schnurr1997, Levine_Lubensky2001a, Norris2006} and in the
presence of compression~\cite{Levine_Lubensky2001a}.   In this work we include the effects of
sliding between the sphere and the medium in a
two-fluid model for a gel with phases representing a viscoelastic
polymer network and fluid solvent.  Throughout this paper, we work in
the dilute network limit, which is appropriate for many
microrheological studies, such as those of actin
networks~\cite{Levine_Lubensky2001a} and DNA solutions~\cite{Chen2003}.
Sliding is implemented using a
kinetic friction law---the shear stress of the network is proportional to the
relative velocity between the network and the surface of the sphere,
with proportionality constant $\Xi$.  No-slip ($\Xi =\infty$) and
frictionless sliding ($\Xi = 0$) boundary conditions can be obtained
as limiting cases.

Physically, we can see the
effects of slip by comparing two different situations:  a solid sphere
of radius $a$ moving with velocity $\mathbf{v}$ through a liquid of viscosity $\eta$, and a clean
bubble of radius $a$ moving through the same liquid.  In the first
case, there is no slip at the surface of the sphere, and the drag
force is $\mathbf{f}= -6 \pi \eta \mathbf{v} a$.  In the second case,
the drag on the bubble is  smaller, $\mathbf{f}= -4 \pi \eta \mathbf{v} a$.
In the far field, the velocity field for the fluid flow for the bubble
is also reduced by a factor of $2/3$ relative to flow for the solid sphere~\cite{batchelor1967}.   

Our use of the two-fluid model for a gel allows us to include the effects of sliding via a boundary
condition between the sphere and polymer network, even as the solvent
retains no-slip boundary conditions.
Because the fluid and network
are coupled by drag, far from the sphere there is no relative motion
between the network and the solvent.   
However, because the fluid and network have
different boundary conditions, the far-field motion can have the
character of that driven by no-slip boundary conditions, that driven
by frictionless sliding boundary conditions, or somewhere in between.
We find that the far-field motion is controlled by frequency.  In the
high frequency limit, the far-field solution has the properties of a
deformation driven by no-slip boundary conditions.  In this case, the
solvent flow is the same as the flow around a sphere in a simple
viscous fluid
and no-slip boundary conditions, and the network is dragged along by
the solvent.  
In the low frequency limit,
the far-field solution has the properties of a deformation driven by
frictionless 
sliding boundary conditions.  The network displacement is the same as
the displacement around a sphere in a simple elastic solid and
frictionless boundary conditions, and the drag prevents the solvent
from moving faster than the network.

Similarly, the force felt by the sphere
interpolates between the limits provided by no-slip and
frictionless sliding boundary conditions.   
The effects of sliding and compression can be described using
    three crossover frequencies.  In the following, $a$ is the radius
    of the probe particle, $l$ is the mesh size of the polymer
    network, $\eta$ is the viscosity of the solvent, $\mu$ and
    $\lambda$ are effective Lam\'e coefficients of the network
    (Eq. \ref{gelequations1}), and $\Xi$ is the friction coefficient
    between the probe particle surface and the polymer network
    (Eq. \ref{fullBC}). 
\begin{enumerate}
\item Associated with sliding, $\omega_s \approx |\mu| l/(\eta a)$ for
  $a>l$, and $\omega_s \approx |\mu| l^2/(\eta a^2)$ for
  $a<l$.
\item Associated with sliding friction, $\omega_f \approx |\mu|/(\Xi
  a)$.
\item Associated with compression, $\omega_c \approx |2 \mu
  + \lambda| l^2/ (\eta a^2 )$.
\end{enumerate}
The properties of the
force can be summarized as
\begin{enumerate}
\item[(a)] At frequencies larger than $\omega_s$, $\omega_f$, and
  $\omega_c$, the response force obeys Eq.~\ref{rheo}, and moduli
  deduced using the generalized
  Stokes-Einstein relation match macroscopic measurements.
\item[(b)] At frequencies smaller than $\omega_s$ and $\omega_f$, but
  larger than
  $\omega_c$, effects from sliding are important, the response force is reduced, and moduli
  deduced using the generalized
  Stokes-Einstein relation underestimate by up to 33\%.
\item[(c)] At frequencies smaller than $\omega_c$, but larger than
  $\omega_s$ or $\omega_f$, effects from compression are important,
  the response force is reduced, and moduli
  deduced using the generalized
  Stokes-Einstein relation underestimate by up to 20\%.
\item[(d)] At frequencies smaller than $\omega_s$, $\omega_f$, and
  $\omega_c$, effects from both sliding and compression are important,
  the response force is reduced, and moduli
  deduced using the generalized
  Stokes-Einstein relation underestimate by up to 43\%.
\item[(e)] In real systems, $\mu$ and $\lambda$ may be frequency dependent,
  and therefore so are $\omega_{s,f,c}(\omega)$.  In light of the above, if at
  any point in an experiment $\omega \lesssim \omega_{s,f,c}(\omega)$, 
  caution should be exercised in applying the generalized
  Stokes-Einstein relation.  Our
  results provide a way to analyze microrheological data in this
  more complicated situation.
\end{enumerate}

The paper is organized as follows.  In Section~\ref{model} we
introduce the two-fluid model for a gel and discuss how sliding
boundary conditions can lead to qualitatively new behaviors.  We
discuss these behaviors in the simple geometry of a gel between two
oscillating plates.  In Section~\ref{solution} we describe the
analytic solution for the force exerted on an oscillating sphere in a
gel, with some details
relegated to Appendix~\ref{solutiondetails}.  Then we describe the
properties of the force in Section~\ref{incompressible}.  All the
features associated with sliding are present in the case of an
incompressible network, which we discuss in the main body of the
paper.  In the discussion we
describe the implications for microrheological experiments, providing
examples of how our results can be used to help interpret
microrheological data.  Finally, in Appendix~\ref{compressible} we 
describe the properties of the force in the presence of both sliding
and a compressible network for completeness.  

\section{The importance of boundary conditions in the two-fluid model for a gel}\label{model}

\begin{figure}
\includegraphics[width=4.5cm]{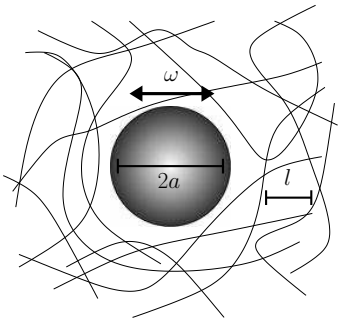}
\caption{A probe particle modeled as a sphere of radius $a$
  oscillating with frequency $\omega$ in a gel modeled by two phases
  representing a viscous solvent and a viscoelastic polymer network
  with mesh size $l$.}\label{spherefig}
\end{figure}

We describe the gel as a polymer network with Lam\'e coefficients $\mu$
and $\lambda$, interspersed with fluid with viscosity $\eta$ (Fig.~\ref{spherefig}).  In
general, we assume that the network shear modulus is complex, and
write $\mu = \mu_1 - i \mu_2$.  
The polymer and fluid are coupled to each other by a friction
coefficient $\Gamma$~\cite{Levine_Lubensky2001a, deGennes1976a,
  Milner1993}.   
We ignore any inertial contributions to the equations of motion:
\begin{eqnarray}
\mu \nabla^2 \mathbf{u} + (\lambda + \mu) \nabla(\nabla \cdot \mathbf{u}) &=& \Gamma \left( \dot{
\mathbf{u}} - \mathbf{v} \right) \label{gelequations1}\\
\eta \nabla^2 \mathbf{v} - \nabla p &=& -\Gamma \left( \dot{\mathbf{u}} - \mathbf{v} \right) \label{gelequations2}\\
\nabla \cdot \left[ (1 - \varphi) \mathbf{v} + \varphi \mathbf{u}\right]&=&0.\label{gelequations3}
\end{eqnarray}
Here $\mathbf{u}$ is the displacement field of the polymer network and
$\mathbf{v}$ is the velocity field of the solvent.  $\varphi$ is the
volume fraction of the polymer network, and 
Eq.~\ref{gelequations3} expresses the overall mass conservation of the gel.  In the limit of small volume
fraction, we can approximate Eq.~\ref{gelequations3} as the incompressibility of the
solvent, $\nabla \cdot \mathbf{v} =0$.  In Eqs.~\ref{gelequations1}
and \ref{gelequations2}, the dot denotes a material time derivative; 
throughout this
paper we replace this with a partial time derivative since
we only consider small displacements and work with these equations
only to linear order.  To deduce the forces exerted by the gel, we use the
stress tensors
\begin{eqnarray}
\bm{\sigma}^{\mathrm{poly}} &=& \mu \left[ \nabla \mathbf u
+ (\nabla \mathbf{u})^\mathrm{T} \right] + \mathbf{I} \; \lambda \nabla \cdot \mathbf{u}  \\
\bm{\tau}^{\mathrm{fluid}} &=& \eta \left[ \nabla \mathbf{v} + (\nabla \mathbf{v}
)^\mathrm{T} \right] \\
\bm{\sigma}^{\mathrm{fluid}} &=& \bm{\tau}^{\mathrm{fluid}} - p \;
  \mathbf{I} .
\end{eqnarray}
It is important to note that the moduli are effective moduli for the
network in the presence of the solvent including, for example, osmotic
effects. 
In addition, $\mu$ and
$\lambda$ have implicit dependence on volume fraction $\phi$ since
the network stiffness depends on the network density.  For a typical
microrheological experiment, we are interested in a dilute network,
with mesh size much larger than the network filament diameter.  For
such a dilute network, $\Gamma\approx \eta/l^2$, where $l$ is the mesh
size of the network~\cite{Levine_Lubensky2001a}.   We note that
Eqs.~\ref{gelequations1} and \ref{gelequations2} are not restricted to
the dilute network limit; for example, in a dense network, these
equations reduce to the poroelastic equations if $\bm{\tau}^{\mathrm{fluid}}$
is negligible compared to the other stresses in the system~\cite{BarryHolmes2001}.  

This model has previously been used by 
Levine and Lubensky~\cite{Levine_Lubensky2001a} to confirm that even
with compressional effects, the force on a sphere obeys Eq.~\ref{rheo} for a range of frequencies.  
Above this range of frequencies
inertial effects become important, while below this range of
frequencies, the compressibility of the material can become
important.  So far, the effects of
compressibility have not been observed~\cite{MacKintosh_Levine2008}. 

In this work, we also use the two-fluid model, so we can
incorporate the compressibility effects described above.  We ignore
inertial effects, which typically are not important at frequencies used in microrheological
measurements.  
However, while previous results were obtained for no-slip boundary
conditions, we obtain exact analytical results
using boundary conditions which allow the sphere to slide with respect
to the polymer network.  Our boundary conditions encompass both the
no-slip and frictionless sliding limits with respect to the polymer network.  

Although Norris~\cite{Norris2006} and Oestreicher~\cite{Oestreicher1951} have
previously obtained analytic solutions for the response function of a
sphere in a material with a single complex shear modulus for both no-slip and
frictionless sliding boundary conditions, a crucial feature of the two-fluid model is that the boundary conditions for the fluid and the
network can be different.  This feature opens up the possibility that the
far-field behavior can \trp{reflect} 
either the fluid or the
network boundary conditions, depending on the frequency of motion.

\begin{figure}
\includegraphics[width=8.5cm]{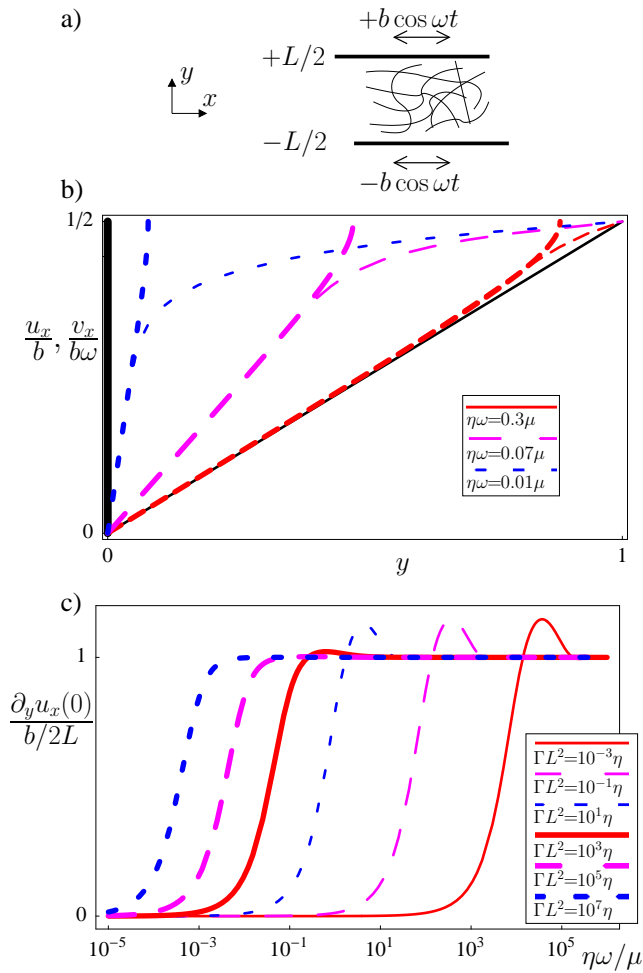}
\caption{(color online) a) Dimensionless amplitude of oscillations as
  a function of dimensionless height for solvent
  velocity (thin lines) and polymer network
  displacement (thick lines) in a gel between two oscillating plates.  The
  polymer network has frictionless boundary conditions at the plate surface.
The dashed lines have $\Gamma L^2/\eta = 10^{2.5}$ and  $L \ll d$.
  Far away from the plate the network and solvent move together.
  The solid line has  $\Gamma L^2/\eta = 10^{-2.5}$, $\eta
  \omega/\mu = 0.01$, and $d \gg L$.  In this case the network and solvent
  need not move together as a unit far from the plate.
b)  Semilogarithmic plot of the dimensionless slope of the amplitude
  of oscillations of the network displacement at $y=0$ as a function
  of dimensionless frequency.  The shape and spacing of the plots indicates that the
  crossover frequency $\omega_\mathrm{plate}$ behaves as $\Gamma^{-1/2}$ for $\Gamma L^2/ \eta >1$, and as $\Gamma^{-1}$ for $\Gamma L^2/ \eta <1$.}\label{plates}
\end{figure}

A simple demonstration of this frequency-dependent behavior
can be seen in a gel between two oscillating plates located at $y= \pm L/2$
(Fig. \ref{plates}a).    The plates
oscillate with amplitude $\pm b \cos \omega t$ along the $\hat{\bf{x}}$
direction, with no-slip boundary
conditions between the plates and the solvent, $v_x(y=\pm L/2) = \mp b\omega
\sin \omega t$.  First consider no-slip boundary conditions between
the plates and the network, $u_x(y=\pm L/2) = \pm b\omega
\sin \omega t$~\cite{Levine_Lubensky2001a}.  In this case
there is no relative motion between the network and the solvent, and both
undergo simple shear motion \trp{for all frequencies.} 
\trp{Since the strain is homogeneous,}
the stress
exerted on upper plate is by definition determined by the macroscopic
shear modulus $G$.  \trp{Writing} 
this stress as $\sigma_{xy} = \mathrm{Re}[
  \tilde \sigma \exp(-i\omega t)]$ \trp{leads to}
\begin{equation}
G = - \frac{\tilde \sigma}{b/(2L)} = \mu_1 - i (\mu_2 + \eta\omega).
\end{equation}

When the network can slip past the plate, the situation becomes more complicated, \trp{since the strain need not be homogeneous}.  For
example, suppose there is
frictionless sliding between the plates and the polymer
network, \trp{so that the tangential stress  on the network vanishes at the plates, $\partial_y u_x(y=\pm L/2) = 0$.}
For
simplicity assume $\mu$ is real ($\mu_2=0$).  Only the $x$-components of the velocity
and displacement fields are nonzero, and the solutions are
\begin{widetext}
\begin{eqnarray}
v_x &=&  -\mathrm{Re} \left\lbrace \frac{2 i b \omega
  \left(\frac{\eta\omega k}{i \mu} y \cosh{\frac{k L}{2
  }}    + \sinh{k y} \right) }{\frac{k L \eta
  \omega}{i \mu}  \cosh{\frac{k L}{2}} + 2
  \sinh{\frac{k L}{2}}} \exp{(-i
  \omega t)}\right\rbrace,\\
u_x &=&  -\mathrm{Re} \left\lbrace \frac{2 i b \omega
  \left(\frac{\eta\omega k}{ \mu} y \cosh{\frac{k L}{2
  }}   - \frac{\eta \omega}{\mu}  \sinh{k y}\right) }{\frac{k L \eta
  \omega}{i \mu}  \cosh{\frac{k L}{2}} + 2
  \sinh{\frac{k L}{2}}} \exp{(-i
  \omega t)}\right\rbrace.
\end{eqnarray}
\end{widetext}
In these equations we introduce the complex quantity  
$k = \sqrt{
  \Gamma (1/\eta - i \omega /\mu)}$
which implies an associated length scale, 
\begin{equation}
d = \frac{1}{\mathrm{Re}\left\lbrace \sqrt{
  \Gamma (1/\eta - i \omega /\mu)} \right\rbrace}.
\end{equation}
\trp{The penetration depth $d$ determines the thickness of the layer next to each plate where the solvent moves relative to the network. Friction suppresses relative motion beyond the penetration depth.} 

\trp{The velocity and displacement fields are plotted in
  Fig.~\ref{plates}b. To see the range of behavior that is possible,
  it is useful to consider limiting cases. First, suppose there is no
  coupling between the network and the solvent, $\Gamma=0$. Then the
  network has zero displacement and the solvent has homogeneous
  oscillatory strain (similar to solid black lines,
  Fig.~\ref{plates}). Now suppose $\Gamma\neq0$. At sufficiently high
  frequency, the penetration depth
  $d\approx\sqrt{2\mu/(\Gamma\omega)}$ is small compared to $L$, and
  there is no relative motion between the network and the solvent in
  most of the gel. Deep within the gel, the solvent has a constant
  strain rate and the network has a constant strain. The penetration
  depth $d$ determines the size of the boundary layers where the
  displacement changes from uniform strain to no strain to satisfy the
  condition of zero tangential stress at each plate. On the other
  hand, when the frequency is sufficiently low, $d$ is approximately
  the mesh size $l$ and we may still have $L\gg d$ if $\Gamma$ is big enough. In this case, again there is no relative motion between the solvent and the network deep in the gel, but now the deformation of the network is small. The penetration depth determines the thickness of the boundary layers at each plate where the velocity of the solvent rises from the small velocity of the network to the velocity of the plate.}  
  

\trp{To determine the crossover frequency that marks the transition
  between the high- and low-frequency behavior, it is convenient to
  plot the strain of the network at the midplane of the gel,
  $\partial_y u_x(0)$, as a function of dimensionless frequency,
  $\eta\omega/\mu$ (Fig.~\ref{plates}c) . At high frequency, we see
  that $\partial_y u_x(0) \approx b/2 L$ -- the network is dragged
  along with the solvent and the motion in the interior of the gel is
  simple shear. At low frequency, $\partial_y u_x(0) \approx 0$ -- the deformation of the network and velocity of the solvent in the interior of the gel are small. Figure~\ref{plates}c shows the crossover for various values of the coupling $\Gamma$. The scaling of the crossover frequency $\omega_\mathrm{plate}$ with $\Gamma$ can be deduced by examining the spacing of the curves in Fig~\ref{plates}c.}
  \trp{F}or $\Gamma>\eta/L^2$, the crossover frequency is approximately $\omega_\mathrm{plate} \approx
  \mu /(L \sqrt{\Gamma \eta} )$, with power law $\Gamma^{-1/2}$; while for
  $\Gamma < \eta/L^2$, the crossover frequency is approximately
  $\omega_\mathrm{plate} \approx \mu/(L^2 \Gamma)$, with power law $\Gamma^{-1}$.
  Later we will see similar behavior of the crossover frequency as a
  function of $\Gamma$ for the spherical geometry.

\begin{figure}
\includegraphics[width=4.5cm]{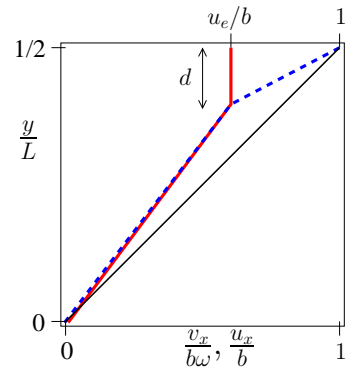}
\caption{(color online) Simplified profile of dimensionless oscillation amplitudes for solvent
  velocity (blue dashed) and
  network displacement (red solid) for estimating force balance.  The
  thin black line is the profile for a single fluid between the plates.} \label{plateestimate}
\end{figure}

\begin{figure}
\includegraphics[width=8.5cm]{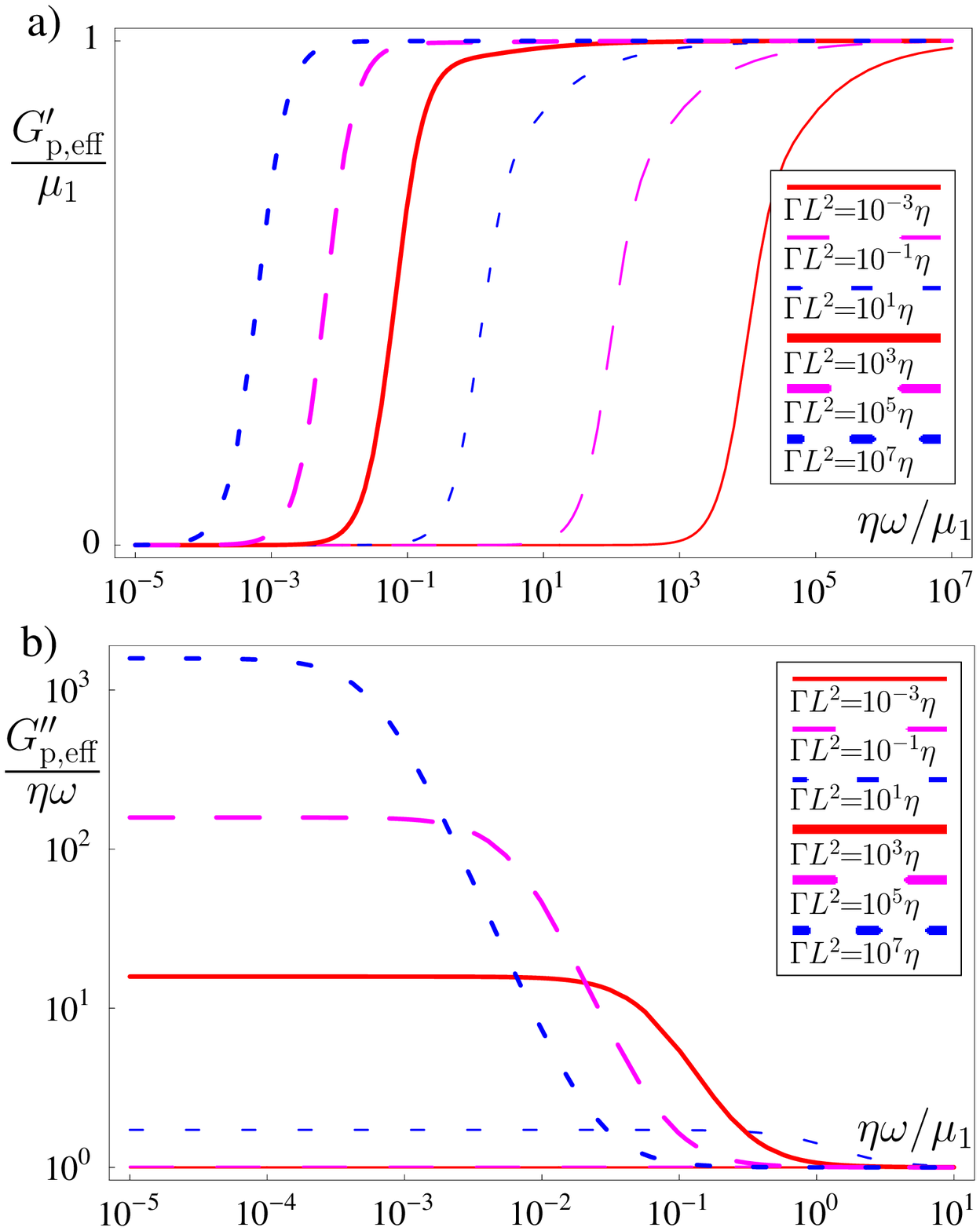}
\caption{(color online) a) Semilogarithmic plot of dimensionless
  $G'_{\mathrm{p,eff}}$ determined by the stress exerted by the gel on
  the plates, as a function of dimensionless frequency.   
b)  Logarithmic plot of $G''_{\mathrm{p,eff}}/(\eta\omega)$ as
  a function of dimensionless frequency.
  $G''_{\mathrm{p,eff}}$ is scaled by the frequency so the plotted quantity is
  proportional to an apparent viscosity.     
}\label{plateforces}
\end{figure}

\trp{When $\Gamma>\eta/L^2$, we observe from Fig.~\ref{plates}c that} 
the crossover frequency satisfies $\eta
\omega_\mathrm{plate} /\mu \ll 1$, and \trp{thus conclude that} 
$d \approx \sqrt{\eta/\Gamma}$ \trp{for $\omega$ near $\omega_\mathrm{plate}$. Let us further suppose that $d\ll L$.}  In this
case we can 
understand the crossover frequency by examining force balance on the
gel.   \trp{As emphasized earlier,} 
the solvent and network move together \trp{except in the boundary layers near the plates}.  The
plates exert no direct force on the network, so any shear on the
network 
results from the motion of the solvent dragging the network along in
the boundary layer, and any stress from network shear in the interior
is ultimately supported by the traction of the solvent at the plate.   
The forces
can be estimated 
\trp{by approximating $v_x$ and $u_x$ with piecewise linear functions, as}
in Fig. \ref{plateestimate}.  In the limit of zero Reynold's number,
there is force balance on any given layer of gel. 
Consider 
\trp{a} layer bounded by the plate on one side, with width $d$.  At the
plate, the traction is of magnitude $\eta\omega(b-u_e)/d$, \trp{where $u_e$ is the displacement at distance $d$ from the plate}.  At the other side of
the layer, the solvent and network stresses add up to approximately
$(\eta \omega +\mu) u_e/(L/2 - d)$.    Equating these two \trp{stresses} 
\trp{determines} $u_e$. 
\trp{Using} $\partial_y
u_x(0) \approx u_e/(L/2- d)$, \trp{we find that}
the normalized strain $\partial_x u(0)/[b/(2L)]\approx(1/[1+ 2 \mu
d/(\eta L \omega) ]$.   For $\omega > \mu d/(\eta L) \approx \mu /(L \sqrt{\Gamma \eta} )$ 
\trp{the normalized strain} is
approximately one, meaning that $u \approx b$, while for $\omega < \mu
/(L \sqrt{\Gamma \eta} )$ the 
\trp{normalized strain} decreases rapidly.  
\trp{Therefore, our force balance argument yields the same crossover frequency we identified from Fig.~\ref{plates}c.}
Physically, the traction due to the network shear
is supported by the drag force between the network and solvent
in the boundary layer of relative motion.  
The higher the frequency, the faster the relative
motion, the more drag force, and the greater the network shear will be
in the interior of the gel. 

\trp{This} crossover behavior \trp{also} shows up in the stress exerted by the gel
on the plates.   Suppose the force on the plates is interpreted
as a measurement of the macroscopic shear modulus.  
Again expressing the stress exerted on the upper
plate as $\sigma_{xy} = \mathrm{Re}[ \tilde \sigma
  \exp(-i\omega t)]$, we obtain
\begin{eqnarray}
G'_{\mathrm{p,eff}} = - \frac{\mathrm{Re}\; \tilde\sigma}{b/(2L)}\\
G''_{\mathrm{p,eff}} = \frac{\mathrm{Im}\; \tilde\sigma}{b/(2L)}.
\end{eqnarray} 
Due to the slip and relative motion, the gel does not undergo a
homogeneous shear deformation, and $G_{\mathrm{p,eff}}$ need not be
equal to $G$. 

At high
  frequencies, the network is pulled by drag to move like the
  uncoupled solvent, and $G'_{\mathrm{p,eff}}\approx\mu_1$
  (Fig.~\ref{plateforces}a), showing the response of an elastic solid
  being sheared between the plates.  Below the crossover frequency,
  $G'_{\mathrm{p,eff}}$ decreases rapidly, taking the character of a network
  with frictionless sliding boundary conditions, which does not exert
  stress on the plates.
Similarly, at high frequencies, $G''_{\mathrm{p,eff}}\approx\eta\omega$
(Fig.~\ref{plateforces}b), showing the viscous response of a simple
fluid with viscosity $\eta$ sheared between the two plates.  In Fig.~\ref{plateforces}b, the
quantity plotted is $G''_{\mathrm{p,eff}}/\eta\omega$, which is
proportional to an apparent viscosity of the medium.  Below the
crossover frequency and for $L\gg d$, the apparent viscosity is
enhanced.  The profile of
Fig.~\ref{plateestimate} explains this enhancement:  at low frequencies, $u_e$ is nearly zero,
and the solvent shear is approximately $b\omega/ d$.  Therefore the
viscous stress on the plate is approximately $\eta b\omega/ d \sim
\Gamma^{1/2}$.  
In the rest of this paper we describe analogous physics which arises in
the spherical geometry, and discuss its ramifications for
microrheological experiments.

\section{Solution for sphere moving in gel}\label{solution}

Consider a sphere of radius $a$ surrounded by a gel.  
We solve for the flows and displacements in the frame of the sphere,
using spherical coordinates $(r,\theta,\phi)$.  We consider a sphere oscillating along the
$\hat {\mathbf{z}}$ direction ($\theta = 0$) with amplitude $B
\cos(\omega t)$.

In the frame of the sphere, the boundary conditions for the
fluid and solid at $r= \infty$ are $\mathbf{v}(\infty,\theta,\phi) = -
( \cos \theta,-\sin \theta ,0) B \omega \sin (\omega t)$
and $\mathbf{u}(\infty, \theta,\phi) = (\cos \theta,-\sin \theta,0 )
B \cos (\omega t)$.  
The
boundary conditions at the surface of the sphere are
\begin{eqnarray}
\mathbf{v}(a,\theta,\phi) &=& 0 \nonumber\\
u_r(a,\theta,\phi) &=& 0 \nonumber\\
\sigma^{\mathrm{polymer}}_{r\theta}(a,\theta,\phi)  &=& \Xi \dot{u}_\theta(a,\theta,\phi). \label{fullBC}
\end{eqnarray}
Here the dot denotes a time derivative,  and
$\mathbf{\sigma}^\mathrm{polymer}$ is the stress tensor of the polymer
network.  The last condition allows the polymer network to slide past
the surface of the sphere in the tangential direction, with a stress
exerted proportional to the relative velocity.  The magnitude of the
force is controlled by the friction coefficient $\Xi$.  For $\Xi
\rightarrow \infty$ we obtain no-slip boundary conditions, while
for $\Xi =0$ we obtain frictionless sliding.

In dimensionless form, Eqs. \ref{gelequations1}-\ref{gelequations3} and Eq. \ref{fullBC} are
\begin{eqnarray}
\frac{1}{R} \mathrm{curl}^2 \mathbf{u} - \frac{1}{\rho}
\nabla (\nabla \cdot \mathbf{u}) &=& -\gamma (\dot{\mathbf{u}} -
\mathbf{v})  \label{nondimgelequations1}\\
\mathrm{curl}^2 \mathbf{v} + \nabla P &=& \gamma (\dot{\mathbf{u}} -
\mathbf{v})\label{nondimgelequations2}
\end{eqnarray}
\begin{eqnarray}
\mathbf{v}(\infty,\theta,\phi) &=& - (\cos \theta,-\sin \theta  ,0) b
\sin t \nonumber
\\
\mathbf{u}(\infty, \theta,\phi) &=& ( \cos \theta,-\sin \theta ,0 ) b
\cos t \nonumber \\
\mathbf{v}(a,\theta\phi) &=& 0 \nonumber\\
u_r(a,\theta,\phi) &=& 0 \nonumber\\
\sigma^\mathrm{polymer}_{r\theta}(a,\theta,\phi)  &=& \xi
\dot{u}_\theta (a,\theta,\phi), \label{nondimBC}
\end{eqnarray}
where we measure lengths in units of $a$, time in units of $1/\omega$, and
pressures and stresses in units of $\eta \omega$.  The dimensionless
parameters are $R = \eta \omega/\mu$, $\rho = \eta \omega/ (2 \mu +
\lambda)$, $\gamma = \Gamma a^2/\eta$, $b = B/a$, and $\xi = \Xi a/\eta $.
As mentioned before, $\Gamma \approx \eta/ l^2$~\cite{Levine_Lubensky2001a}, where $l$ is the
mesh size of the polymer network, so that $\gamma \approx a^2/l^2$.   
For convenience we keep the same notation for dimensionless
fields as we used for dimensional fields.

Our method is an
extension of the classic solution by Stokes of a sphere moving through
a viscous fluid~\cite{batchelor1967}. 
The velocity field is divergence-free and can be expressed in terms of
a single stream function $\Psi$:
\begin{equation}
\mathbf{v} = \mathrm{curl} \left( \frac{\Psi \hat {\bm{\phi}}}{r \sin
  \theta} \right).
\end{equation}
On the other hand, the polymer is compressible, so the displacement
field has both a stream function for the divergence-free part ($\Phi$),
and a potential function for the curl-free part ($\chi$):
\begin{equation}
\mathbf{u} = \mathrm{curl} \left( \frac{\Phi \hat {\bm{\phi}}}{r \sin \theta}\right) +
\nabla \chi.
\label{displacementstream}
\end{equation}

Taking the curl and divergence of the above equations, leads to
\begin{eqnarray}
-\frac{1}{\rho} \nabla^4 \chi &=& - \gamma \, \nabla^2 \dot \chi \label{chieq}\\
\frac{1}{R} \mathrm{curl}^4
\left(\frac{\Phi \hat {\bm{\phi}}}{r \sin \theta}\right) &=& -\gamma \, \mathrm{curl}^2 \left( \frac{\dot \Phi \hat {\bm{\phi}}}{r \sin \theta}
-\frac{\Psi \hat {\bm{\phi}}}{r \sin \theta} \right) \label{phieq}\\
\mathrm{curl}^4
\left(\frac{\Psi \hat {\bm{\phi}}}{r \sin \theta}\right) &=& \gamma \,  \mathrm{curl}^2 \left( \frac{\dot \Phi \hat {\bm{\phi}}}{r \sin \theta}
-\frac{\Psi \hat {\bm{\phi}}}{r \sin \theta} \right) \label{psieq}.
\end{eqnarray}
The solution for $\chi$ is written in the appendix, and involves three
undetermined coefficients.

Diagonalizing the equations for $\Phi$ and $\Psi$ yields
\begin{equation}
\mathrm{curl}^4 \left(
\begin{array}{c} \alpha \\ \beta \end{array}\right) \frac{\hat
  {\bm{\phi}}}{r \sin \theta} = \mathrm{curl}^2  \left(
\begin{array}{cc} \lambda_\alpha & \\ &\lambda_\beta \end{array}\right) \left(
\begin{array}{c} \alpha \\ \beta \end{array}\right) \frac{\hat
  {\bm{\phi}}}{r \sin \theta} \label{diagonalized} 
\end{equation}
where
\begin{equation}
 \left(
\begin{array}{c} \alpha \\ \beta \end{array}\right)
= \mathbf{M}  \left(
\begin{array}{c} \Phi \\ \Psi \end{array}\right),
\end{equation} 
and $\mathbf{M}$
is a two-by-two matrix.  The
appendix contains expressions for the matrix $\mathbf{M}$ and eigenvalues
$\lambda_\alpha$, $\lambda_\beta$ which
diagonalize the equations.  The appendix also contains the solutions for
$\alpha$ and $\beta$; each involves
three undetermined coefficients.

The nine undetermined coefficients are determined by imposing the
boundary conditions at infinity and the sphere surface.  In addition,
one must demand that Eq. \ref{nondimgelequations1} is
satisfied by the combination of stream and potential functions in
Eq. \ref{displacementstream}.  The conditions on the coefficients are
displayed in the appendix and determine the velocity and displacement fields.

From the velocity and the displacement fields the (nondimensional)
stresses and pressure of the polymer and fluid can be calculated:
\begin{eqnarray}
\bm{\sigma}^{\mathrm{poly}} &=& \frac{1}{R} \left[ \nabla \mathbf u
+ (\nabla \mathbf{u})^\mathrm{T} \right] + \mathbf{I} \; \left(\frac{1}{\rho} -
\frac{2}{R} \right) \nabla \cdot \mathbf{u}  \\
\bm{\tau}^{\mathrm{fluid}} &=& \nabla \mathbf{v} + (\nabla \mathbf{v}
)^\mathrm{T} \\
p &=& \int \mathrm{d}r\;\left[\gamma (\dot u_r - v_r) + \nabla^2 v_r\right]. 
\end{eqnarray}
The total stress tensor is 
\begin{equation}
\bm{\sigma}^{\mathrm{total}} = \bm{\sigma}^\mathrm{poly} + \mathbf{\tau}^{\mathrm{fluid}} - p \;\mathbf{I}. 
\end{equation}
Integrating the stress tensor over the surface of the sphere yields
the total force on the sphere, which is (in dimensionless form)
\begin{widetext}
\begin{eqnarray}
\mathbf{f} &=& \mathrm{Re} \left\lbrace f e^{-i\omega t}  \right\rbrace\\
f &=& -  6 b \pi \frac{i R \xi [2i +
    (\rho + 2 R) + (1- i R) Y] - [4 i + (2 \rho + 6 R) +(2-3 i R) Y'
+ (1- i R) Z] }{i R \xi [ i(\rho + 2 R)
    + R Y] -  [i(2 \rho + 6 R) +3 R Y' + R Z] }
    \label{force}\\
Y &=&  i 2 \sqrt{-i \gamma \rho} + \gamma \rho + \frac{ \sqrt{\gamma}} {
    \sqrt{1 - i R}} \rho \nonumber \\
Y' &=&  i 2 \sqrt{-i \gamma \rho} \nonumber \\
Z &=&  \frac{\sqrt{\gamma}}{\sqrt{1 - i R}} \left( 2 \rho + 2 R -i R
    Y'  \right) . \nonumber
\end{eqnarray}
\end{widetext}
In these expressions, whenever a square root appears, the root with
positive real part should be chosen.  This expression is one of the
main results of this paper.  Although it is in analytic form, it is
still unwieldy.  In the next section we summarize its properties. 

\section{Properties of response force}\label{incompressible}

The dimensional response force is Eq. \ref{force} multiplied by
$a^2 \eta \omega$.  If interpreted using the generalized Stokes-Einstein
relation, this yields effective moduli 
\begin{eqnarray}
G'_{\mathrm{eff}} &=& -\frac{\eta \omega}{6 \pi} \;\mathrm{Re} \left\lbrace \frac{f}{B
  e^{-i\omega t}} \right\rbrace\\
G''_{\mathrm{eff}} &=& \frac{\eta \omega}{6 \pi} \; \mathrm{Im}\left\lbrace \frac{f}{B
  e^{-i\omega t}} \right\rbrace.
\end{eqnarray}
If Eq.~\ref{rheo} holds, we would expect
$G'_{\mathrm{eff}} = \mu_1$, and $G''_{\mathrm{eff}} = \mu_2 + \omega \eta$.
In the following we will describe the response force in terms of the
effective moduli.   The descriptions in the following sections are
obtained by analyzing the form of Eq. \ref{force}.  
To give clarity
to the discussion, in all the plots we assume that $\eta$ and $\mu$ are
frequency-independent.     

To understand the properties of the response force when the particle
can slide past the network, first we look at
the limit of incompressible network, $\lambda \rightarrow \infty$, which
 displays all of the effects due to sliding boundary conditions.
Network compressibility (finite $\lambda$) complicates the situation by adding additional
features to the effective moduli.  We leave a detailed description of the more complicated
scenario with effects from both sliding and compressibility to the
appendix, but summarize the results in the discussion.  

\subsection{Frictionless limit with incompressible network}

To illustrate the effects of sliding when the network is incompressible, 
we concentrate on the case of frictionless
sliding between the sphere and the polymer network, $\xi =0$.    
When both the network and solvent are incompressible and have no-slip
boundary conditions, there is no relative motion between the network
and the solvent, and Eq.~\ref{rheo} holds
~\cite{Schnurr1997}.  In
contrast, when the polymer network has frictionless boundary
conditions at the surface of the sphere, the network and solvent move
relative to each other near the sphere.  This relative motion between the network
and solvent leads to drag forces which tend to diminish the relative
motion, so that far from the sphere the network and solvent move
together.   The far field solution can be 
either that of an incompressible material driven by no-slip boundary
conditions, that of an incompressible material driven by
frictionless boundary conditions, or somewhere in between.  As in the
example of a gel between two oscillating plates,  
at high frequencies, the solution driven by the velocities (the
solvent, with no-slip boundary conditions) wins out, while at low
frequencies the solution driven by displacements (the polymer network,
with frictionless boundary conditions) wins out.  

In the limit of frictionless sliding ($\xi \rightarrow 0$) and
incompressibility ($\rho \rightarrow 0$) Eq.~\ref{force} reduces to
\begin{equation}
f_{\mathrm{inc}} = -6 b \pi \frac{i (2 - 3 i R) + (1-iR) \frac{R
    \sqrt{\gamma}}{\sqrt{1- i R}}}{R \left[ 3 i + \frac{R
    \sqrt{\gamma}}{\sqrt{1- i R}}\right] }.   
\end{equation}
The low frequency limit is obtained when the first term in the
brackets in the denominator dominates over the second term, and the
high frequency limit is obtained in the opposite case.  Therefore the 
crossover frequency $\omega_s$ satisfies $|R
    \sqrt{\gamma}/\sqrt{1- i R}| \approx 1$.  For $\gamma \approx
    a^2/l^2 >1$ (sphere larger than mesh size),
    $\omega_s \approx |\mu|/(\eta \sqrt{\gamma})$, while for $\gamma <1$
    (mesh size larger than sphere), $\omega_s \approx
    |\mu|/(\eta\gamma)$.  
The expression for $\omega_s$ in the introduction is written in terms
of variables which can be easily controlled experimentally using the
approximation $\gamma \approx
    a^2/l^2$, and hence applies when $\Gamma\approx \eta/l^2$.
Note that in these two limits, the form of
    the crossover is the same as in the case of the plates but with
    the length scale $L$ replaced by $a$; in particular the dependence
    on $\Gamma$ is the same as in the case of the oscillating plates,
    with power law $\Gamma^{-1/2}$ and $\Gamma^{-1}$, respectively.

\begin{figure}
\includegraphics[width=8.5cm]{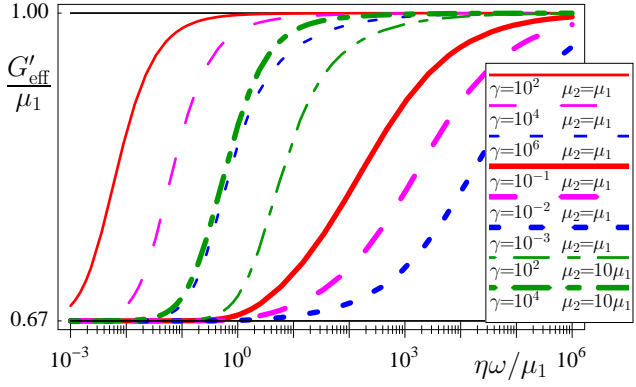}
\caption{(color online) Logarithmic plot of dimensionless
  $G'_{\mathrm{eff}}$ as a function of dimensionless
  frequency with frictionless boundary conditions between sphere and
  an incompressible polymer network.  Eq.~\ref{rheo} would predict
  $G'_{\mathrm{eff}} = \mu_1$.   The shape and spacing of the plots
  indicates that there is a crossover frequency $\omega_s
  \approx |\mu|/(\eta \sqrt{\gamma})$ for $\gamma>1$, and $\omega_s
  \approx |\mu|/(\eta \gamma)$ for $\gamma<1$.  For $\omega<\omega_s$,
  effects from sliding cause $G'_{\mathrm{eff}}$ to underestimate $\mu_1$
  by up to 33\%.} \label{incompressibleG'}
\end{figure}

For frequencies  $\omega \gg \omega_s$, the modulus $G_{\mathrm{eff}}$ obeys Eq.~\ref{rheo}, so that $G'_{\mathrm{eff}} = \mu_1$ and
$G''_{\mathrm{eff}} = - (\eta \omega + \mu_2)$.  For 
$\omega \ll \omega_s$, $G_{\mathrm{eff}}$ behaves as
\begin{eqnarray}
G_{\mathrm{eff}}(\omega\ll\omega_s) &=& \left(\frac{2}{3} \mu_1 - \frac{2}{9} \frac{
  \sqrt{\gamma} \mu_2 \mu_1 \eta \omega}{\mu_1^2 + \mu_2^2} \right) \\
&&- i \left( \frac{2}{3} \mu_2 + \eta\omega \left[1 + \frac{\sqrt{\gamma}
  (\mu_1^2 - \mu_2^2)}{9( \mu_1^2 + \mu_2^2)} \right]\right). \nonumber
\end{eqnarray}
For the lowest frequencies, the real modulus $G'_{\mathrm{eff}} = 2
\mu_1/3$, 67\% of the value in Eq.~\ref{rheo}.
The behavior of $G'_{\mathrm{eff}}$ as a function of frequency is plotted in 
Fig. \ref{incompressibleG'}.

\begin{figure}
\includegraphics[width=8.5cm]{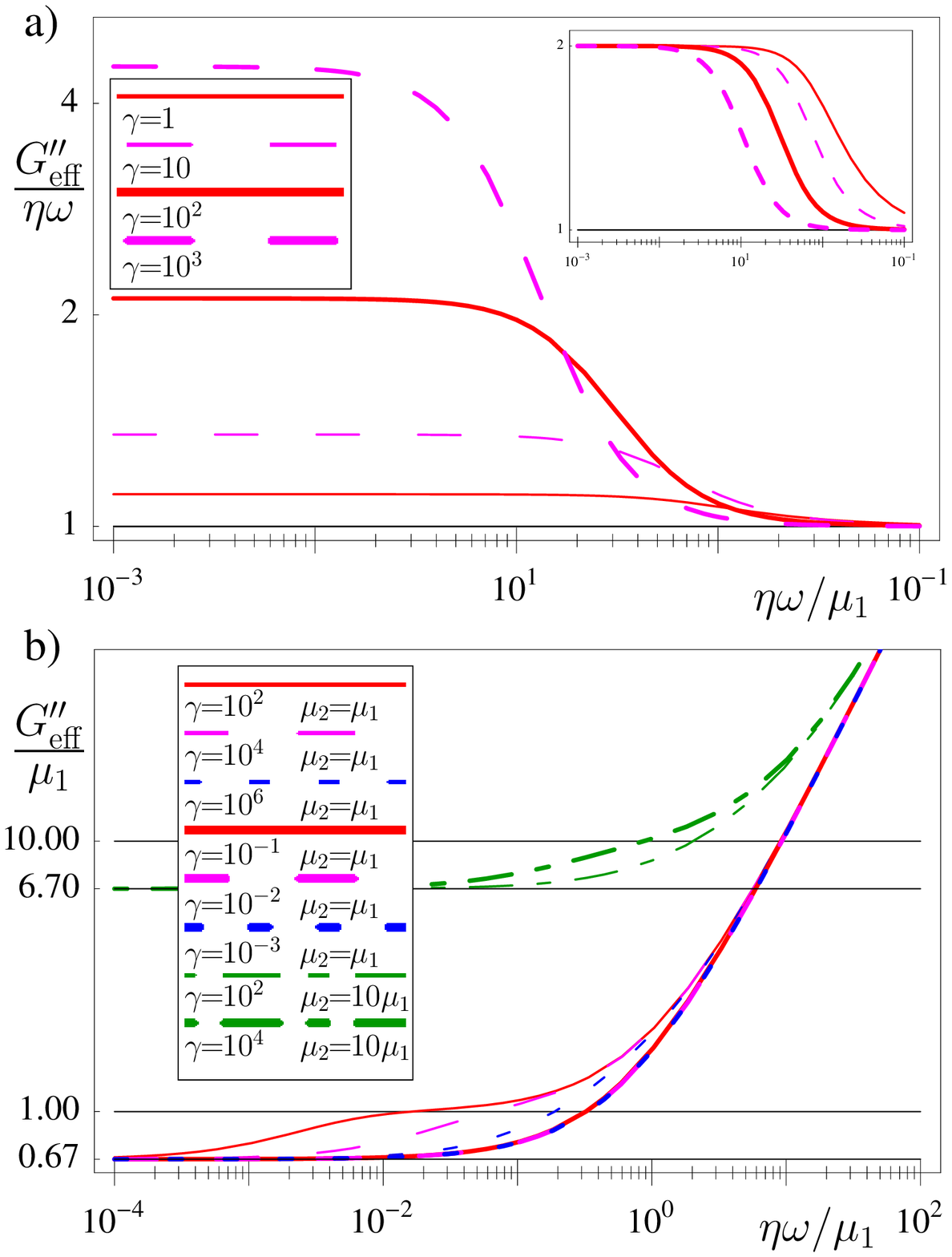}
\caption{(color online) Logarithmic plots of dimensionless
  $G''_{\mathrm{eff}}$ as a function of dimensionless 
  frequency for frictionless boundary conditions between the sphere
  and an incompressible polymer network.
a) $G''_{\mathrm{eff}}/(\eta \omega)$ as a function of frequency for $\mu_2
  =0$.  This contribution to $G''_{\mathrm{eff}}$ from solvent viscosity and
  network-solvent drag dominates for small $\mu_2$ or high
  frequencies.   Inset:  $G''_{\mathrm{eff}}$ rescaled by its low frequency
  magnitude (see text) to demonstrate that the crossover frequency is
  $\omega_s$.      b)  $G''_{\mathrm{eff}}$ as a function of
  frequency.  Eq.~\ref{rheo} would predict
  $G''_{\mathrm{eff}} = \mu_2 + \eta \omega$.  For $\omega<\omega_s$,
  effects from sliding cause $G''_{\mathrm{eff}}$ to underestimate $\mu_2$
  by up to 33\%.  The crossover behavior is the same as that of the
  real part ($G'_{\mathrm{eff}}$) except that for $\omega>\mu_2/\eta$
  the solvent viscosity ($\eta \omega$) dominates.
} \label{incompressibleG''}
\end{figure}

Now we turn to the imaginary modulus $G''_{\mathrm{eff}}$.
Eq.~\ref{rheo} predicts that $G''_{\mathrm{eff}} = \mu_2 + \eta
\omega$.  In the example of the plates, we assumed $\mu_2=0$, and 
found an enhancement
in the effective viscosity at low frequencies.  In the case of a
sphere there is a similar enhancement to the effective viscosity, but
sliding also affects the estimation of $\mu_2$.    

First consider the case where the network shear modulus is purely
real ($\mu_2 =0$).  In this case, all dissipation arises from the
viscous solvent  and friction between the solvent and the polymer
network.  
The behavior of $G''_{\mathrm{eff}}$ with $\mu_2 =0$ is shown in
Fig. \ref{incompressibleG''}a, $G''_{\mathrm{eff}}/(\eta \omega)$, an
apparent viscosity.
When $\omega \gg \omega_s$, the apparent viscosity is $\eta$.
For frequencies $\omega \ll \omega_s$, the
apparent viscosity of the ``solvent'' contribution tends towards a
plateau with enhanced viscosity $\approx \eta (1 +
\sqrt{\gamma}/9)$.  
Physically, as in the case of the plates, the
solvent moves more than the network in a layer near the sphere.  A
region of enhanced solvent shear and viscous drag is set up as the
solvent reduces its motion relative to the network, until far away the
solvent takes the same motion as the network.  As in the case of the
plates, the enhancement depends on $\Gamma$ as $\Gamma^{1/2}$. 
In the inset to Fig. \ref{incompressibleG''}a   we plot $G''_{\mathrm{eff}}$ rescaled by
  its low frequency value, $1+ (G''_{\mathrm{eff}} - \eta
  \omega)/ (\eta\omega \sqrt{\gamma}/9)$, to show that
  the crossover value is $\omega_s$.  

Now consider the case when the imaginary part of the network modulus
($\mu_2$) is nonzero.    In addition to the contribution to
$G''_{\mathrm{eff}}$ described in the previous paragraph, there are
also contributions from the network modulus $\mu_2$.
For  $\omega \gg \omega_s$, the leading contribution is $\mu_2$, the
result of Eq.~\ref{rheo}.  When $\omega \ll
\omega_s$, the leading contribution is $2 \mu_2/3$, 66\% of the
high-frequency contribution.   For a constant $\mu_2$, at high frequencies the contribution
proportional to $\eta \omega$ dominates, while at low frequencies the
contribution from $\mu_2$ dominates.  In total, $G''_{\mathrm{eff}}$
behaves as shown in Fig. \ref{incompressibleG''}b.  

\subsection{Incompressible network with intermediate friction}

\begin{figure}
\includegraphics[width=8.5cm]{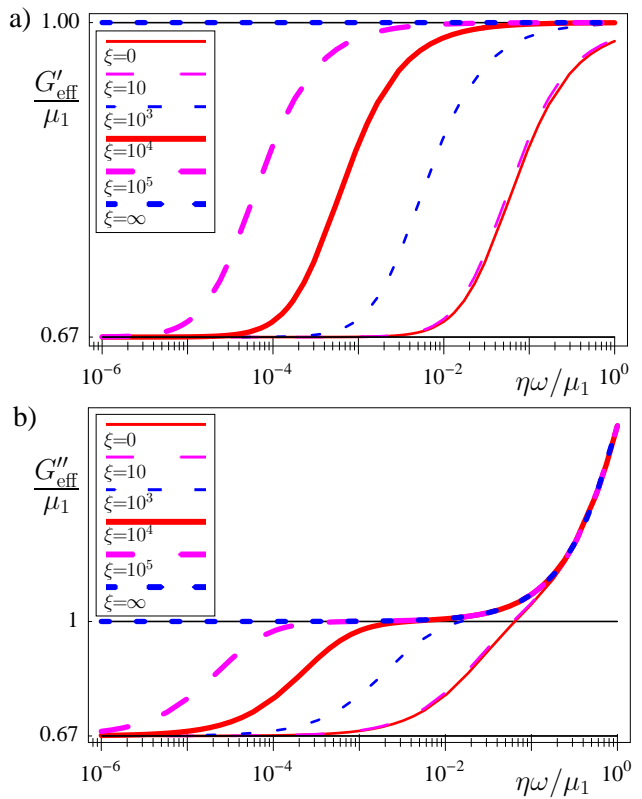}
\caption{(color online) Logarithmic plot of dimensionless a)
  $G'_{\mathrm{eff}}$, b) $G''_{\mathrm{eff}}$ as a function of dimensionless
  frequency for an incompressible network and for boundary conditions
  with intermediate friction between sphere and polymer network.  In these plots $\mu_1=\mu_2$ and
  $\gamma=10^4$.  The behavior interpolates between that
  of no-slip boundary conditions at high frequencies and frictionless
  boundary conditions at low frequencies, with crossover frequency
  $\omega_f \approx |\mu|/(\Xi a)$ .  For frequencies below both
  $\omega_s$ and $\omega_f$ the moduli can be
  underestimated by up to 33\%.} \label{fullGincompressible}
\end{figure}

For any finite value of $\xi$, there will be a crossover from behavior
similar to $\xi = 0$ at low frequencies to behavior similar to $\xi
\rightarrow \infty$ at high frequencies, as shown in Fig. \ref{fullGincompressible}.  
The crossover is set by the condition $|R|\xi=1$ and occurs at $\omega_f \approx \left| \frac{ \mu}{\Xi a}
\right|$.  Because the no-slip limit of $G'_{\mathrm{eff}}$ is  $\mu_1$,
$G'_{\mathrm{eff}}$ appears to crossover from a low frequency limit of $2 \mu_1
/3$ to the high frequency limit $\mu_1$ at $\omega = \mathrm{min}\lbrace
\omega_f, \omega_s\rbrace$.  Similar behavior is seen in the imaginary
modulus $G''_{\mathrm{eff}}$.

\section{Discussion}\label{discussion}

The main aim of this paper is to provide an understanding of how sliding
between a probe particle and polymer network can affect the response
function in a gel.  Motivated by experimental data, Starrs and
Bartlett~\cite{Starrs_Bartlett2003} have previously suggested the
possiblity of frictionless sliding boundary
conditions being relevant in microrheological experiments.  Our
solution shows how this may come about, despite the fact that the liquid solvent always couples to probe
particles with no-slip boundary conditions in continuum fluid
mechanics.  We find that, due to
the coupling between the solvent and polymer network, in  far-field,
the solvent can move as if it is driven by sliding boundary conditions.  Specifically, when the fluid and polymer network have different boundary
conditions, at high frequencies this coupling tends to lock the network
into the motion of the fluid, while at low frequencies the coupling
locks the fluid into the motion of the network.  

Our analytical
solution for the force on a probe particle in response to oscillatory
motion quantifies the effect of sliding.
We have described the effects of sliding on the force felt by probe
particles in terms of an effective modulus $G_{\mathrm{eff}}$.  For
high enough frequencies, $G_{\mathrm{eff}}$ matches the amcroscopic
shear modulus $G$, while for low frequencies $G_{\mathrm{eff}}$ may
underestimate $G$.  We have identified two crossover frequencies, one
associated with sliding ($\omega_s$), and one associated with friction
($\omega_f$).  For any finite amount of surface friction, when
frequency is reduced below both $\omega_s$ and $\omega_f$ ($\omega \ll
\mathrm{min}\lbrace \omega_s, \omega_f\rbrace$), $G_{\mathrm{eff}}$
tends to underestimate moduli by 33\%.  

The description in the previous
paragraph applies in the case of an incompressible network, as treated in
the body of this paper.  However, in real materials the
compressibility of the network can lead to additional reductions in
$G_{\mathrm{eff}}$.  These compressional effects complicate the
behavior of $G_{\mathrm{eff}}$ as a function of frequency, but may be
relevant in experimental situations, so we have described them in detail
in Appendix~\ref{compressible}.  Intuitively, a compressible network
is softer than an incompressible network and should exert a smaller
force on an oscillating sphere.  Our analytic solution identifies
a third crossover frequency  associated with compressive effects
($\omega_c$).  This is the same crossover frequency identified earlier
by other investigators~\cite{Schnurr1997, Levine_Lubensky2001a}.    In
terms of the effective modulus $G_{\mathrm{eff}}$, the effects of
compression and sliding together are as follows:  For frequencies much
greater than $\omega_s$, $\omega_f$, and $\omega_c$,
$G_{\mathrm{eff}}$ matches the macroscopic modulus $G$.  For
frequencies such that $\mathrm{min}\lbrace \omega_s,\omega_f \rbrace
\ll \omega \ll \omega_c$, compressional effects reduce $G_{\mathrm{eff}}$ by up to 20\%.  For frequencies such that
$\omega_c \ll \omega \ll \mathrm{min}\lbrace \omega_s,\omega_f
\rbrace$, sliding effects reduce $G_{\mathrm{eff}}$ by up
to 33\%.  Finally, for the lowest frequencies, much less than
$\omega_s$, $\omega_f$, and $\omega_c$, both compressional and sliding effects reduce $G_{\mathrm{eff}}$ by up to 43\% relative to $G$.

For any specific experiment, our results describe how the appropriate effects of sliding
on single-particle measurements should be accounted for. 
For example, if $\gamma\approx a^2/l^2>1$, which occurs for particles
larger than the mesh size, then a good estimate for
$\omega_s$ is given by $\eta \omega_s/ |\mu|  \approx 1/\sqrt{\gamma} \approx l/a$.  Therefore $\omega_s$ can be
estimated in experimental measurements, allowing determination of
whether it is possible for corrections to Eq.~\ref{rheo} from sliding effects to be
important.  In general, $\mu$ will be frequency dependent, and
therefore so will $\omega_s(\omega)$.  A simple test which follows
from our analysis is that if throughout the range of
experimental frequencies $\omega \gg \omega_s(\omega)$, then one can
determine that sliding will not affect the effective moduli.
In the following two paragraphs, we apply this test to
two sets of experimental results reported in the literature.

Starrs and Bartlett~\cite{Starrs_Bartlett2003} have observed a 2/3 reduction in single-particle
moduli for frequencies above $10\,\mathrm{rad}/\mathrm{s}$ in polystyrene solutions.
Using the solvent viscosity of decalin ($2.6\, \mathrm{mPa}-\mathrm{s}$), a trap spring
constant of $\approx 1.66\times 10^{-6}\,\mathrm{N}/\mathrm{m}$, and 
Fig. 6 and 7 of Ref.~\cite{Starrs_Bartlett2003}, $\eta \omega/|\mu|$ can
be estimated to be 0.3 -- 0.6 for frequencies between $10 \,\mathrm{rad}/\mathrm{s}$ and $10^4\,
\mathrm{rad}/\mathrm{s}$.  To obtain $1/\sqrt{\gamma} \approx l/a$ for  the semidilute
polystyrene solution (concentration $c = 1.7 c^*$), we
use a bead size $a=0.64\,\mu \mathrm{m}$~\cite{Starrs_Bartlett2003}, and  estimate the correlation length as $l = R_g (c^* /c) \approx 0.06
\,\mu\mathrm{m}$~\cite{Brochard_deGennes1977}, where $R_g\approx
100\,\mathrm{nm}$~\cite{Starrs_Bartlett2003} is the radius of gyration.  We obtain
the estimate $1/\sqrt{\gamma} \approx 10^{-1}$.  From these estimates
we conclude that
$\omega$ may be slightly greater than
$\omega_s$, but not by orders of magnitude.  Therefore the effects
of sliding may be important.   

On the other hand, Buchanan {\it et. al.}~\cite{Buchanan2005} have reported that the generalized
Stokes Einstein relation is accurate in wormlike micellar solutions.
Using the solvent viscosity of water, and the moduli from Fig. 1 of
Ref.~\cite{Buchanan2005}, it can be estimated that 
$\eta \omega/|\mu|$ is between $10^{-3}$ and $10^{0}$ for
frequencies where the one particle microrheology data is reported, and in most
cases $\eta \omega/|\mu| > 10^{-2}$.  Using a mesh lengthscale of
$10\,\mathrm{nm}$ and particle radius of $1\,\mu\mathrm{m}$,
$1/\sqrt{\gamma} \approx l/a \approx 10^{-2}$.  Our calculation would then lead to the conclusion that for most of the data
reported, and certainly for frequencies $\nu > 100\, \mathrm{Hz}$ [for which
$\eta \omega/|\mu|$ is at least an order of magnitude greater than
$10^{-2}$], the effects of sliding should not be important.  In this
system, discrepancies from the generalized Stokes-Einstein relation due to
sliding effects can only appear for frequencies below  $\approx 1\,\mathrm{Hz}$.

In the examples provided above, we have focused on
estimating $\omega_s$, rather than $\omega_f$.  However, if $\omega_f<\omega_s$, sliding effects will not appear until
$\omega < \omega_f$.   This dependence on friction may provide a way
to test the predictions of our analysis by varying the surface
chemistry of the probe particles~\cite{McGrath2000}.

While in most of this discussion we have focused on probe particles larger
than the mesh size ($\gamma\approx a^2/l^2 >1$), our results also
apply when the probe particle is smaller than the mesh size ($\gamma <
1$).   In this
case, $\omega_s \approx |\mu|/(\eta \gamma)$ can be large and may be
more easily accessible to experiments.    In addition, when probe
particles are much smaller than mesh size, they may be more likely to
not directly interact with the network, leading to small values of the
friction coefficient and large values for $\omega_f$.  Our solution provides a framework to analyze
experiments in this regime of small probe particle size. 

Finally, an additional observation from the form of our results is that there is the
possibility for features of the frequency dependence of moduli, including
power laws, to be obscured or distorted in the vicinity of the crossover
frequencies $\omega_s$, $\omega_c$, and $\omega_f$.  This suggests that
caution may be in order near these crossover frequencies.

In the literature there has been much discussion of how probe-material
interactions can affect one-particle microrheological measurements.
For example, near a probe particle the polymer network can be depleted
by steric hindrance.   One effect of this depletion zone is to change
local rheological properties, but another may be to increase the
likelihood that the bead slides relative to the polymer
network~\cite{Starrs_Bartlett2003}.  Our calculation takes into account the
second effect but not the first.  Quantitative comparison of
$G_{\mathrm{eff}}$ to experiments may require treatment of both
effects, as well as the effects from adhesion of probe surface to polymer
networks~\cite{Valentine2004}.

\textbf{Acknowledgements}  We thank C. Wolgemuth for helpful
discussions.  This work was supported in part by National
Science Foundation grants DMS-0615919 (TRP) and CMMI-0825185 (VS), and
a Solomon Faculty Research Grant from Brown University (VS).  TRP
thanks the Aspen Center for Physics, where some of this work was completed.   

\appendix
\section{Details of solution}\label{solutiondetails}

The solution to the equation for the stream function $\chi$ can be
obtained by using the axisymmetry of the solution, so that $\chi$ is
independent of $\phi$.  In addition, due to the boundary conditions at
infinity, $\chi$ must be
proportional to $\cos \theta$.  Using these conditions in
Eq. \ref{phieq}, we find that the radial
dependence of $\chi$ satisfies an ordinary differential equation.  The
solution to $\chi$ which results in finite displacements at infinity is
\begin{equation}
\chi = \mathrm{Re} \left\lbrace \left[ \frac{G}{r^2} + H r + J (\frac{1}{r^2} +
\frac{k_\chi}{r} ) e^{- k_\chi r} \right] e^{-i t} \cos\theta \right\rbrace, 
\label{chi}
\end{equation}
where $k_\chi = \sqrt{-i \gamma \rho}$, in which the root with
positive real part is chosen.

To obtain Eq. \ref{diagonalized} we need to find the matrix
$\mathbf{M}$ which diagonalizes the equations for the stream function,
\begin{equation}
\left( \begin{array}{cc} \lambda_\alpha & \\ & \lambda_\beta
\end{array}\right) = \mathrm{M} \left( \begin{array}{cc} i \gamma R &
  \gamma R \\ -i \gamma & -\gamma \end{array} \right) \mathrm{M}^{-1}.
\end{equation}
In the above time derivatives have been evaluated with respect to the
oscillatory function $\exp(-i t)$ which multiplies all parts of the
solutions.  Explicitly, $\mathrm{M}$ is 
\begin{equation}
\mathrm{M} = \frac{1}{i + R}\left( \begin{array}{cc}1&R\\-1&i \end{array} \right),
\end{equation}
and the eigenvalues are $\lambda_\alpha = 0$, $\lambda_\beta = -(1-
iR) \gamma$.  

The solutions to $\alpha$ and $\beta$ (Eq. \ref{diagonalized}) are found by a
similar procedure as the solution to $\chi$.  $\alpha$ and $\beta$ must be proportional to $\sin^2\theta$ to satisfy the
boundary conditions. The radial dependence obeys an ordinary
differential equation, and the
solutions which result in finite velocities and displacements at
infinity are 
\begin{widetext}
\begin{eqnarray}
\alpha &=& \mathrm{Re} \left\lbrace \left( A r^2 + \frac{B}{r} + C r
\right) \sin^2 \theta  \; e^{-i t} \right\rbrace \\
\beta &=& \mathrm{Re} \left\lbrace \left( D r^2 + \frac{E}{r} + F(
\frac{1}{r} + \sqrt{-\lambda_\beta}) e^{- \sqrt{-\lambda_\beta} r}  \right) \sin^2 \theta  \; e^{-i t} \right\rbrace. 
\label{alphabeta}
\end{eqnarray}
\end{widetext}
Again, the square roots denote the root with positive real part.  Note
that it would not be difficult to generalize this method of solution
to incorporate the effects of inertial terms.

The boundary conditions Eq. \ref{nondimBC} at $r=a$ yield four (linear)
equations for the coefficients $A-G$, since there both the velocity
and displacement fields each have an $r$ and $\theta$ component.

Additional conditions on the coefficients $A-G$ arise from the boundary
conditions for the displacement and velocity field at $r=\infty$. 
Athough each have an $r$ and a $\theta$ component, the $r$ and
$\theta$ component give identical equations for the coefficients
$A-G$, leaving only 2 independent conditions out of the four:
\begin{eqnarray}
2 M^{-1}_{11} A + 2 M^{-1}_{12} D + H &=& b \\
2 M^{-1}_{21} A + 2 M^{-1}_{22} D &=& - i b.
\end{eqnarray}
In these equations the matrix elements of $\mathbf{M}^{-1}$ appear.

Eq. \ref{nondimgelequations1} has terms proportional to
$\exp(-k_\chi r)$, $\exp(-\sqrt{- \lambda_\beta} r)$, $r^{0}$, and
$r^{-3}$.  The parts proportional to the exponentials are
automatically satisfied by the general solutions in Eqs. \ref{chi} and \ref{alphabeta}.   The terms proportional to $r^0$ and
$r^{-3}$ give nontrivial relations for the coefficients $A-G$.  The $r$ and
$\theta$ component give identical equations for the coefficients,
again leaving two independent conditions out of the four:
\begin{widetext}
\begin{eqnarray}
i (2 M^{-1}_{11} A + 2 M^{-1}_{12} D + H) + (2 M^{-1}_{21} A + 2
M^{-1}_{22} D) &=&0\\
i (2 M^{-1}_{11} B + 2 M^{-1}_{12} E -2 G) + (2 M^{-1}_{21} B + 2
M^{-1}_{22} E) &=& \frac{4 M^{-1}_{11}}{\gamma R} C.
\end{eqnarray}
\end{widetext}
However, note that the first of these (from $r^0$) is not independent from the
equations arising from the boundary conditions at $r=\infty$.

Together, these produce only seven linearly independent equations for
the nine coefficients A-J, so
solving them leaves two coefficients undetermined.  However, these undetermined
coefficients only appear when the stream and potential functions are considered
independently of one another; in the physically meaningful
combinations expressing the velocity and displacement field no
undetermined coefficients remain.  Since the solutions are not illuminating,
we do not write them down.

\section{Properties of the response force with compressible network}\label{compressible}

In this appendix we describe the properties of the response force when
effects from compressibility of the network are included.  First we
isolate the effects of compressibility by examining the case of
no-slip boundary conditions between the sphere and the network.  Then
we include effects from sliding of the sphere past the network.

\subsection{No-slip limit with compressible network}

In this section we address the limit of no-slip between the sphere and
the polymer, $\xi \rightarrow \infty$ when $\lambda$ is finite.   
Examining the form of the solutions, one can see that the stream
function $\chi$ corresponding to the compressive motion of the network
involves an exponential $\exp(- r \sqrt{\rho \gamma}/a)$  with length scale
$\mathrm{Re}[ a/\sqrt{\rho \gamma}]$.   The compressive motion dies off at longer
scales due to the coupling to the incompressible fluid.   Associated
with these compressive effects is a crossover frequency $\omega_c$ set by the
condition $|\rho| \gamma =1$, or $\omega_c \approx |2
\mu + \lambda |/( \eta \gamma)$.  The expression for $\omega_c$ in the
introduction is written in terms of variables which can be easily controlled
experimentally using the approximation $\gamma \approx
a^2/l^2$, and hence applies when $\Gamma\approx \eta/l^2$.
 For frequencies less than $\omega_c$,
compressive effects of the network are important, while for
frequencies above $\omega_c$, compressive effects of the network are
not important.   This crossover is the same as that identified
in previous studies \cite{Schnurr1997,Levine_Lubensky2001a}.

\begin{figure}
\includegraphics[width=8.5cm]{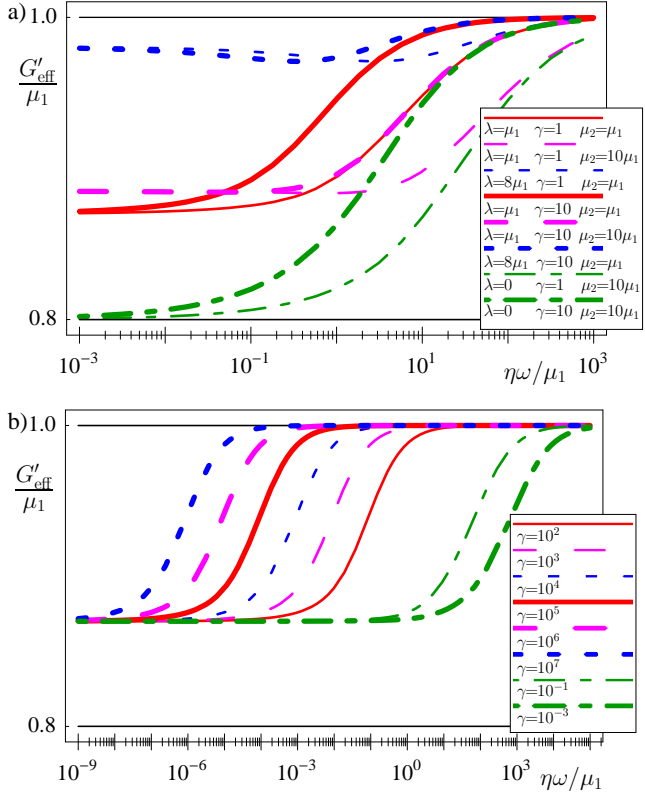}
\caption{(color online) Logarithmic plots of dimensionless
  $G'_{\mathrm{eff}}$ as a function of dimensionless
  frequency for no-slip boundary conditions between sphere and polymer
  network.
Eq.~\ref{rheo} would predict $G'_{\mathrm{eff}}=\mu_1$.  The shape of
  the plots indicates a crossover frequency $\omega_c \approx |2
  \mu+\lambda|/(\eta \gamma)$.  For  the effective
  modulus crosses over to new behavior in which compressional modes of
  the network are important.  At low frequencies G'$\omega<\omega_c$
  effects from compression cause $G'_{\mathrm{eff}}$ to underestimate
  $\mu_1$ by up to 20\%.   b) $G'_{\mathrm{eff}}$ for a range of
  $\gamma$ and $\lambda = \mu_1 =\mu_2$. } \label{noslipG'}
\end{figure}

For
frequencies $\omega \gg \omega_c$, the real part of the modulus $G'_{\mathrm{eff}}$
tends to $\mu_1$, in accord with the expectations of Eq.~\ref{rheo}.  For frequencies $\omega\ll \omega_c$, the real
modulus is diminished, and for very low frequencies tends to the value
\begin{equation}
G'_{\mathrm{eff}}(\omega \rightarrow 0, \xi \rightarrow \infty) = \mu_1 \frac{
     1 + \frac{\mu_1 (1 + ( \frac{\mu_2}{\mu_1} )^2)}{2(\lambda+2
     \mu_1)} + (\frac{2 \mu_2}{(\lambda +
   2 \mu_1)})^2 }{\left( 1 + \frac{\mu_1}{2(\lambda+2 \mu_1)} \right)^2 +
   \left(\frac{5 \mu_2}{2 (\lambda + 2 \mu_1)}\right)^2 }.\label{zeroNoslip}
\end{equation}
Eq.~\ref{zeroNoslip} ranges from 80-100\% of the high frequency limit $\mu_1$.
For an incompressible network ($\lambda \rightarrow \infty$) there is
no reduction, while maximal reduction (20\%) occurs for $\lambda =0$ and $\mu_2
\gg \mu_1$.  For $\mu_1
\approx \mu_2 \approx \lambda$ (reasonable values for an actin network), the reduction is about 14\%.
In Fig. \ref{noslipG'} the behavior of $G'_{\mathrm{eff}}$ is plotted. 

\begin{figure}
\includegraphics[width=8.5cm]{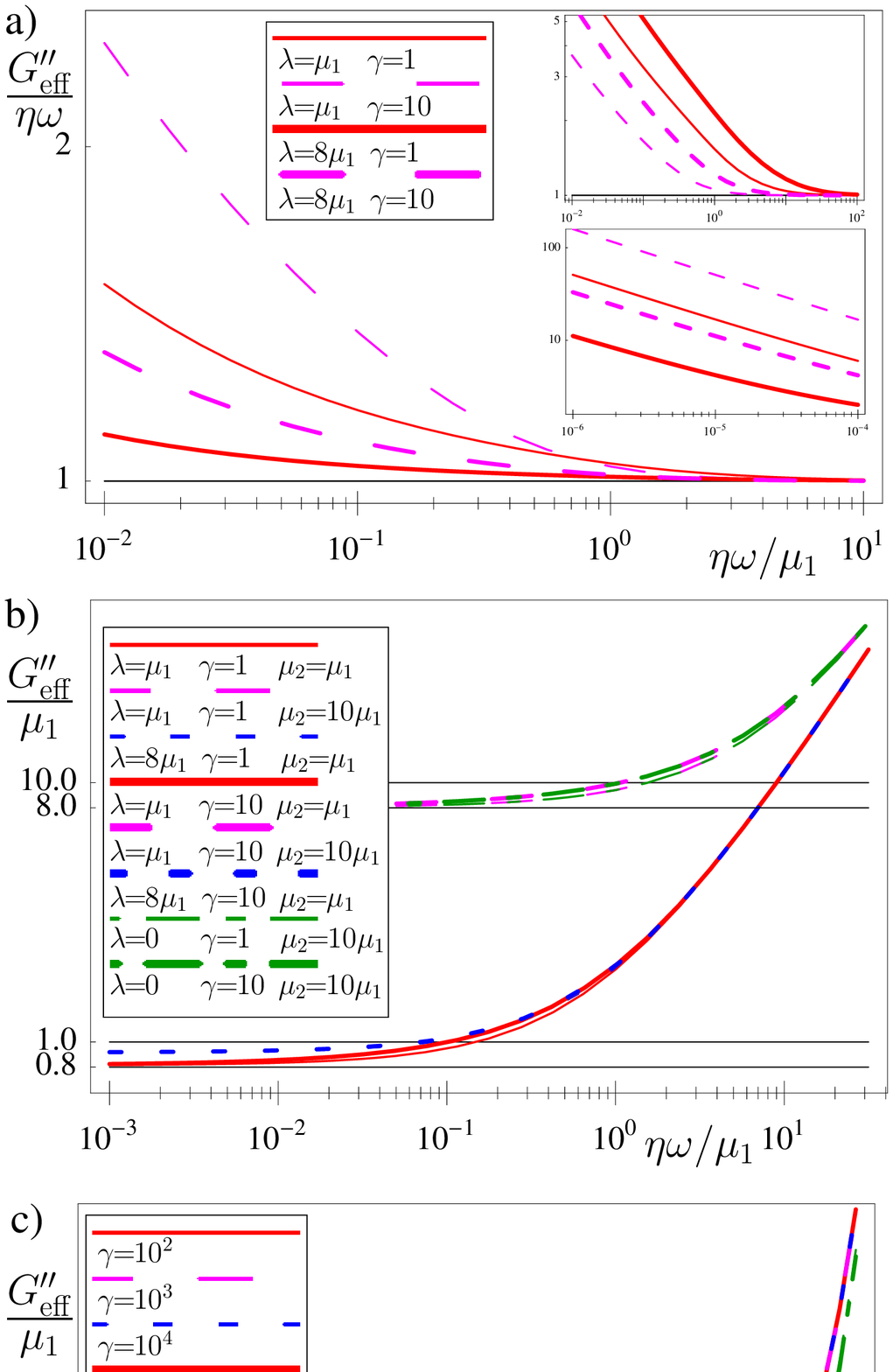}
\caption{(color online) Logarithmic plots of the imaginary part
  dimensionless $G''_{\mathrm{eff}}$ as a function of
  dimensionless frequency for no-slip boundary conditions between
  sphere and polymer network.
a) $G''_{\mathrm{eff}}/(\eta \omega)$ as a function of frequency for $\mu_2
  =0$.  This contribution to $G''_{\mathrm{eff}}$ from solvent viscosity and
  network-solvent drag dominates for small $\mu_2$ or high
  frequencies.   At low frequencies ($\omega<\omega_c$),
  $G''_{\mathrm{eff}} \sim \omega^{1/2}$ (top inset).  Lower inset:
  $G''_{\mathrm{eff}}$ rescaled by its low-frequency magnitude (see
  text) to demonstrate that the crossover frequency is $\omega_c$.
  b)  $G''_{\mathrm{eff}}$ as a function of frequency.  Eq.~\ref{rheo}
  would predict $G''_{\mathrm{eff}}= \mu_2 +\eta\omega$.  For
  $\omega<\omega_c$, effects from compression cause
  $G''_{\mathrm{eff}}$ to  underestimatee $\mu_2$ by up to 20\%.  The
  crossover behavior is the same as in the real part except that for
  $\omega<\mu_2/\eta$ the solvent viscosity ($\eta\omega$) dominates.   
c) $G''_{\mathrm{eff}}$ for a range of $\gamma$ and $\lambda = \mu_1
  =\mu_2$.} \label{noslipG''}
\end{figure}

Now we turn to the imaginary part of the effective modulus $G''_{\mathrm{eff}}$.
As in the incompressible case it is useful to analyze contribution with $\mu_2 = 0$ separately from the polymer network
contribution.  At high frequencies, for $\mu_2=0$, 
$G''_{\mathrm{eff}} = -\omega
\eta$, in agreement with Eq.~\ref{rheo}.
For frequencies below $\omega_c$, $G''_{\mathrm{eff}}$ crosses over to behavior
in which 
\begin{widetext}
\begin{equation}
G''_{\mathrm{eff}}(\omega \ll \omega_c) = \eta \omega \left(1 + \frac{
\sqrt{\frac{2  \gamma \mu_1^4}{\eta \omega 
    (\lambda +2 \mu_1)^3}} + \frac{ (\gamma +
    \sqrt{\gamma}) (\mu_1)^3
   }{ (\lambda+ 2 \mu_1)^3}
  }{\left( \frac{\mu_1}{\lambda+2\mu_1} +2 \right)^2 + 2\left(
  \frac{\mu_1}{\lambda + 2 \mu_1} +2 \right) \sqrt{\frac{2 \gamma \eta
  \omega}{\lambda + 2 \mu_1}}} \right) .  
\label{lownoslipG''} \end{equation}
\end{widetext}
Note that as $\omega \rightarrow 0$, $G''_{\mathrm{eff}} \sim \omega^{1/2}$.  
This behavior is plotted in
  Fig. \ref{noslipG''}a, which shows the quantity $G''_{\mathrm{eff}}/ (\eta
  \omega)$ (an apparent viscosity), so that the high frequency regime appears as a constant,
  while the low frequency regime diverges as $\omega^{-1/2}$.
The low
  frequency divergence of the apparent viscosity
  results from the fact that at low frequencies, the compressive length scale
  $\mathrm{Re}[a/\sqrt{\rho \gamma}]$ diverges, and so there is a larger and larger
  volume with appreciable relative motion between the solvent and network.
  The dissipation at low frequencies is
  increased by the friction between compressive modes of the polymer
  network and the incompressible solvent.   To show that the crossover
  frequency is $\omega_c$, in one of the insets we plot $G''_{\mathrm{eff}}$
  rescaled by its low frequency magnitude, $(G''_{\mathrm{eff}} - \eta
  \omega)/[G''_{\mathrm{eff}}(\omega \ll \omega_c)|_{\omega=\omega_c}
  -\eta \omega] + 1$.

Now consider the case when the imaginary part of the network modulus
($\mu_2$) is nonzero.    In addition to the contribution to
$G''_{\mathrm{eff}}$ described in the previous paragraph, there are
also contributions from the network modulus $\mu_2$.
For high frequencies ($\omega> \omega_c$), this
additional contribution is simply $ \mu_2$, as expected from
Eq.~\ref{rheo}. For low frequencies
$\omega \ll \omega_c$ this contribution is diminished to the limiting value
\begin{equation}
\left. G_{\mathrm{eff}}'' \right|_{
\begin{array}{c} \omega \rightarrow 0\\ \xi \rightarrow \infty \end{array}} = 
     \mu_2 \frac{ (1 + \frac{(\mu_1^2 + 5 \mu_2^2)}{(\lambda +
     2\mu_1)^2} )}{\left( 1 + \frac{\mu_1}{2 (\lambda+2 \mu_1)} \right)^2 +
   \left(\frac{5 \mu_2}{2(\lambda+2\mu_1)}\right)^2 }.
\end{equation}
As in the real part, the low frequency value is between 80\% (for
$\lambda = 0$, $\mu_2 \gg \mu_1$), and 100\% (incompressible, $\lambda =
\infty$) of the high frequency value.  For $\mu_1 \approx
\mu_2 \approx \lambda$, at low frequencies the low frequency
additional contribution is about 81\% of the
high frequency additional contribution.

For a constant $\mu_2$, at high frequencies the contribution
proportional to $\eta \omega$ dominates, while at low frequencies the
contribution from $\mu_2$ dominates.  This yields $G''_{\mathrm{eff}}$ which
behave as shown in Fig. \ref{noslipG''}b and c.  Note that the 
low frequency enhancement discussed in the case $\mu_2 =0$ is not
readily apparent since the contribution from $\mu_2$ dominates at low
frequencies.

\subsection{Frictionless limit with compressible network}

Now we turn to the opposite limit of no friction between the polymer
and sphere ($\xi = 0$).  As in the incompressible case, we expect that
for frequencies below $\omega_s$ the far-field solution takes the
character of a solution with frictionless sliding boundary conditions.
For frequencies  $\omega > \omega_s$, the modulus $G_{\mathrm{eff}}$ obeys Eq.~\ref{rheo}, so that $G'_{\mathrm{eff}} = \mu_1$ and $G''_{\mathrm{eff}} = (\eta \omega + \mu_2)$.  
For frequencies less than $\omega_s$, now that there is also
compressibility, there are two cases to
consider.  In the first case, $\omega_c < \omega_s$, while in the
second case $\omega_c > \omega_s$.    

First, consider the case where
$\omega_c < \omega_s$.   For $\omega_c \ll \omega \ll \omega_s$, the real
modulus $G'_{\mathrm{eff}} = 2 \mu_1/3$, 66\% of the
value in Eq.~\ref{rheo}.  For $\omega \ll \omega_c$,
compressive effects come into play and further diminish $G'_{\mathrm{eff}}$,
which tends to
\begin{equation}
\left. G'_{\mathrm{eff}}\right|_{
\begin{array}{c} \omega \rightarrow 0\\ \xi \rightarrow 0 \end{array}}
 = \frac{2}{3}\mu_1
     \frac{ 1 + \frac{(1 + ( \frac{\mu_2}{\mu_1} )^2) \mu_1}{3
     (\lambda + 2 \mu_1)} + (\frac{2 \mu_2 }{( \lambda + 2\mu_1)})^2 }{\left( 1 + \frac{\mu_1}{3(\lambda+2\mu_1)} \right)^2 +
   \left(\frac{7 \mu_2}{3 (\lambda+2 \mu_1)}\right)^2 }.
\label{slipG'zero}
\end{equation}
The low frequency limit of $G'_{\mathrm{eff}}$ is between 57\% (compressible
limit with $\lambda \rightarrow 0$, $\mu_2 \gg \mu_1$) and 66\%
(incompressible limit, $\lambda \rightarrow \infty$) of the high
frequency value.  For 
$\mu_1 \approx \mu_2 \approx \lambda$, the low frequency limit of
$G'_{\mathrm{eff}}$ is about 60\% of the high frequency value $\mu_1$.

\begin{figure}
\includegraphics[width=8.5cm]{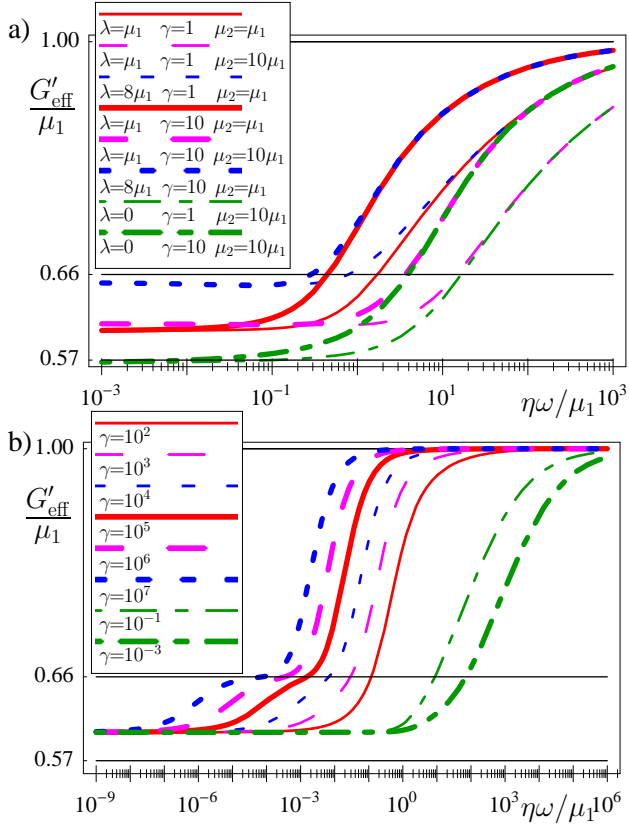}
\caption{(color online) Logarithmic plots of dimensionless
  $G'_{\mathrm{eff}}$ as a function of dimensionless
  frequency for frictionless boundary conditions between sphere and polymer
  network.  Eq.~\ref{rheo} would predict $G'_{\mathrm{eff}} = \mu_1$. 
  a) For frequencies less than both $\omega_s$ and $\omega_c$, effects
  from sliding and compression cause $G'_{\mathrm{eff}}$ to
  underestimate $\mu_1$ by up to 43\%.  b) $G'_{\mathrm{eff}}$ for a
  range of  $\gamma$ and  $\lambda = \mu_1 =\mu_2$.  If
  $\omega_c \ll \omega_s$, for $\omega_c \ll \omega \ll \omega_s$,
  effects from sliding can cause
  $G'_{\mathrm{eff}}$ to underestimate $\mu_1$ by 33\%, and for
  $\omega \ll \omega_c \ll\omega_s$, compressional effects cause an
  additional underestimation by up to 43\% of $\mu_1$.} \label{slipG'}
\end{figure}

In practice the condition that  $\omega_c \ll \omega \ll \omega_s$ may
not be met, either because $\omega_s <\omega_c$, or because
$\omega_c$ is not sufficiently smaller than $\omega_s$.  In this
case $G'_{\mathrm{eff}}$ appears to immediately tend to the
behavior of Eq. \ref{slipG'zero} when $\omega < \omega_s$.   We plot
representative behavior of $G'_{\mathrm{eff}}$ in the frictionless limit in
Fig. \ref{slipG'}.

\begin{figure}
\includegraphics[width=8.5cm]{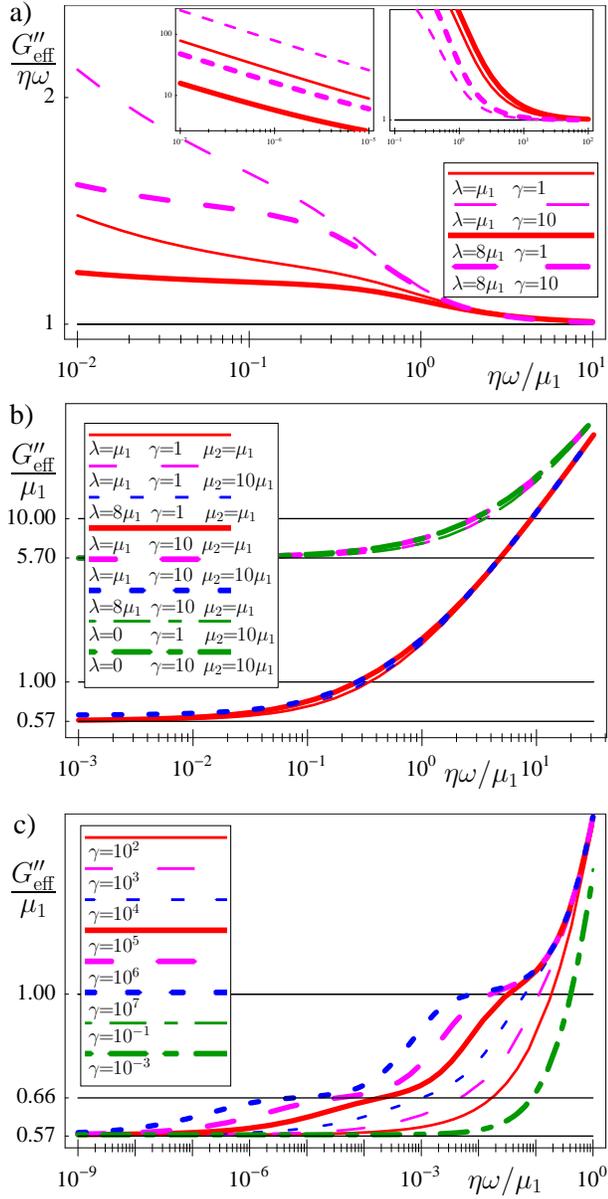}
\caption{(color online) Logarithmic plots of dimensionless
  $G''_{\mathrm{eff}}$ as a function of dimensionless
  frequency for frictionless boundary conditions between sphere and polymer
  network.
a) $G''_{\mathrm{eff}}/(\eta \omega)$ as a function of frequency, for $\mu_2
  =0$.  This contribution to $G''_{\mathrm{eff}}$ from solvent viscosity and
  network-solvent drag dominates for small $\mu_2$ or high
  frequencies.  At the lowest
  frequencies $G''_{\mathrm{eff}} \sim \omega^{1/2}$ (left inset). 
  Right inset: $G''_{\mathrm{eff}}$ rescaled by its low frequency
  magnitude (see text) to demonstrate that the crossover frequency is
  $\omega_s$.      b)  Eq.~\ref{rheo} would predict that
  $G''_{\mathrm{eff}} = \mu_2 + \eta \omega$.  For frequencies less
  than both $\omega_s$ and $\omega_c$, sliding and compressional
  effects cause $G''_{\mathrm{eff}}$ to underestimate $\mu_2$ by up to
  43\%.  The crossover behavior is the same as in the real part except
  that for $\omega>\mu_2/\eta$ the solvent viscosity ($\eta\omega$) dominates. 
c) $G''_{\mathrm{eff}}$ for a range of
  $\gamma$ and $\lambda = \mu_1 =\mu_2$. } \label{slipG''}
\end{figure}

Now we turn to the
imaginary modulus $G''_{\mathrm{eff}}$.  Again, first look at the
contribution arising from the solvent viscosity and drag between the
polymer network and solvent, by setting $\mu_2=0$.  When $\omega \gg \omega_s$, the imaginary modulus $G''_{\mathrm{eff}} = - \eta \omega$.
For frequencies $\omega < \omega_s$, $G''_{\mathrm{eff}}$ crosses over to
behavior of
\begin{widetext}
\begin{equation}
G''_{\mathrm{eff}}(\omega \ll \omega_s, \mu_2 =0) = \eta \omega \left(1 +
\frac{\sqrt{\frac{2 \gamma \mu_1^4}{\eta \omega (\lambda + 2\mu_1)^3}}
  + \frac{\gamma \mu_1^2}{(\lambda+2\mu_1)^2} + \sqrt{\gamma}
  \left(\frac{\mu_1}{\lambda+2 \mu_1} +1 \right)^2  }{
  \left(\frac{\mu_1}{\lambda+2 \mu_1} +3 \right)^2  + 3 \left(
  \frac{\mu_1}{\lambda+2 \mu_1} +3\right) \sqrt{ \frac{2\gamma \eta
      \omega}{\lambda + 2 \mu_1}}
} \right).  
\label{slipsolventG''}
\end{equation}
\end{widetext}
The behavior of $G''_{\mathrm{eff}}$ with $\mu_2 =0$ is shown in
Fig. \ref{slipG''}a, which shows the apparent viscosity
$G''_{\mathrm{eff}}/(\eta \omega)$.
In one of the insets we plot $G''_{\mathrm{eff}}$ rescaled by its low frequency
value, $(G''_{\mathrm{eff}} - \eta
  \omega)/[G''_{\mathrm{eff}} (\omega\ll\omega_s)|_{\omega=\omega_s}
  -\eta \omega] + 1$, to show that the crossover value
  is $\omega_s$.  
Note that for certain parameters, as $\omega$ decreases, one can see
the effective viscosity approach a plateau, which has the same origin
as the plateau in the incompressible case described in the main text.
   At even lower frequencies,
Eq. \ref{slipsolventG''} is dominated by $\omega^{1/2}$ behavior.
This has the same origins in the compressibility of the network as
discussed in the previous section dealing with no-slip boundary conditions.

If $\mu_2 \neq 0$ there is an additional contribution to the imaginary
modulus.  For  $\omega \gg \omega_s$, this contribution is $\mu_2$, the
result of Eq.~\ref{rheo}.  When $\omega_c \ll \omega \ll
\omega_s$, this contribution is $2 \mu_2/3$.  Finally, when $\omega <\omega_c$, this contribution tends
to the limiting value
\begin{widetext}
\begin{equation}
G''_{\mathrm{eff}}(\omega \rightarrow 0, \xi \rightarrow 0) = \frac{2}{3} \mu_2
   \frac{1 + \frac{14
   \mu_2^2 +2 \mu_1^2}{3 (\lambda + 2 \mu_1)^2}}{\left( 1 +
   \frac{\mu_1}{3(\lambda+2 \mu_1)} \right)^2 + \left(\frac{7 \mu_2 }{3 (\lambda+2\mu_1)}\right)^2 }.
\label{slipG''zero}
\end{equation}
\end{widetext}
This limiting value is smaller than the contribution from $\mu_2$ at
high frequencies.  It is between 57\% (compressible limit,
$\lambda=0$, $\mu_2 \gg \mu_1$), and 66\% (incompressible limit,
$\lambda \rightarrow \infty$) of the high frequency contribution
corresponding to Eq.~\ref{rheo}.
For $\mu_1 \approx \mu_2 \approx \lambda$ the limiting value is 58\% of
the high-frequency contribution. 

If $\omega_c > \omega_s$, for $\omega < \omega_s$ the additional
contribution to $G''_{\mathrm{eff}}$ immediately tends to the value in
Eq. \ref{slipG''zero}.

\subsection{Intermediate friction}

For any finite value of $\xi$, there will be a crossover from behavior
similar to $\xi = 0$ at low frequencies to behavior similar to $\xi
\rightarrow \infty$ at high frequencies.  

\begin{figure}
\includegraphics[width=8.5cm]{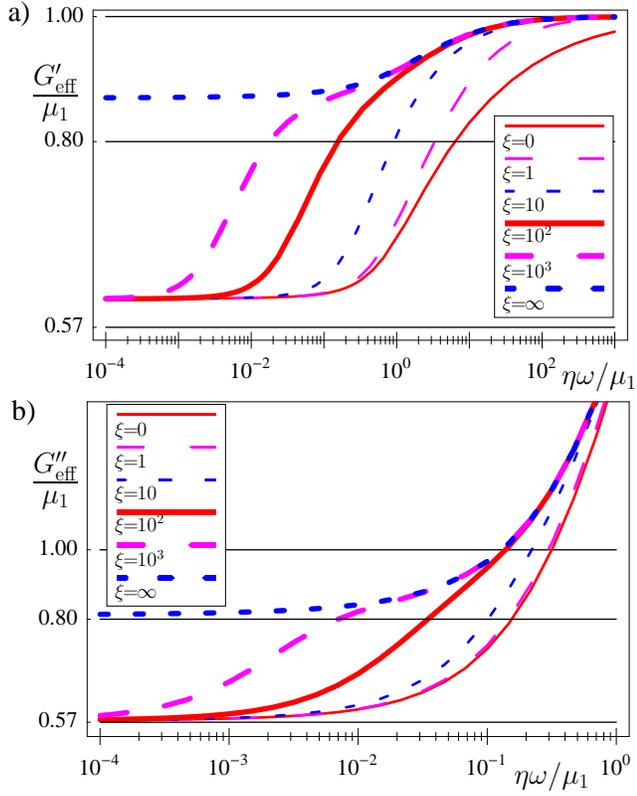}
\caption{(color online) Logarithmic plots of dimensionless a)
  $G'_{\mathrm{eff}}$, b) $G''_{\mathrm{eff}}$ as a function of dimensionless
  frequency for boundary conditions with intermediate friction between
  sphere and polymer network.  In these plots $\lambda=\mu_1=\mu_2$
  and $\gamma = 3$.  The behavior interpolates between that
  of no-slip boundary conditions at high frequencies and frictionless
  boundary conditions at low frequencies, with crossover frequency $\omega_f$.
  For frequencies less than $\omega_s$, $\omega_f$, and $\omega_c$, 
  the moduli can be underestimated by up to 43\%.} \label{fullG}
\end{figure}

The crossover occurs at $\omega_f = \left| \frac{ \mu}{\Xi a}
\right|$.  Typically, as plotted in Fig. \ref{fullG}, 
because the no-slip limit of $G'_{\mathrm{eff}}$ is closer to $\mu_1$,
$G'_{\mathrm{eff}}$ appears to crossover from a low frequency limit of $2 \mu_1
/3$ (or less, if compressional effects are important) to a value
closer to the high frequency limit $\mu_1$ at $\omega = \mathrm{min}\lbrace
\omega_f, \omega_s\rbrace$.  Similar behavior is seen in the imaginary
modulus $G''_{\mathrm{eff}}$, shown in Fig. \ref{fullG}.

\end{document}